%% file: main.tex
\documentclass[12pt]{article}

\RequirePackage[OT1]{fontenc}
\usepackage{amsthm,amsmath,natbib}
\usepackage{array}
\usepackage{float}
\usepackage{booktabs}
\usepackage{tabularx}
\usepackage{xcolor}
\usepackage{color}
\RequirePackage[colorlinks,citecolor=blue,urlcolor=blue]{hyperref}
\usepackage{alltt}
\usepackage{fancyvrb}
\usepackage{algorithm2e}
\usepackage{pdfpages}
\usepackage{multirow}
\usepackage{lineno}
\usepackage{comment}
\usepackage{verbatim}
\usepackage{lscape}
\usepackage{graphicx}

\input{mylatex.tex}

\input{packages.tex}

\def\ie{\textit{i.e.}}
\def\eg{\textit{e.g.}}

\def\i{\textit{(i) }}
\def\ii{\textit{(ii) }}
\def\iii{\textit{(iii) }}
\def\iv{\textit{(iv) }}
\def\v{\textit{(v) }}

\newcommand{\blind}{1}

\begin{document}
\thispagestyle{empty}
\baselineskip=27pt
\vskip 4mm
\begin{center} 

{\Large{\bf Space-Time Landslide Predictive Modelling}}

\end{center}

\baselineskip=12pt
\vskip 3mm

\if1\blind
{
\begin{center}
\large
Luigi~Lombardo$^{1,}$$^{*}$, Thomas~Opitz$^2$, Francesca~Ardizzone$^3$, Fausto~Guzzetti$^3$, Rapha\"{e}l~Huser$^4$
\end{center}

\footnotetext[1]{
\baselineskip=10pt University of Twente, Faculty of Geo-Information Science and Earth Observation (ITC), PO Box 217, Enschede, AE 7500, Netherlands}

\footnotetext[2]{
\baselineskip=10pt INRA, Biostatistics and Spatial Processes, 84914, Avignon, France}

\footnotetext[3]{
\baselineskip=10pt Consiglio Nazionale delle Ricerche (CNR), Istituto di Ricerca per la Protezione Idrogeologica (IRPI), via Madonna Alta 126, 06128 Perugia, Italy}

\footnotetext[4]{
\baselineskip=10pt King Abdullah University of Science and Technology (KAUST), Computer, Electrical and Mathematical Sciences and Engineering (CEMSE) Division, Thuwal 23955-6900, Saudi Arabia}
}
 \fi

\baselineskip=16pt
\newpage

\begin{center}
{\large{\bf Abstract}}
\end{center}
Landslides are nearly ubiquitous phenomena and pose severe threats to people, properties, and the environment in many areas. Investigators have for long attempted to estimate landslide hazard in an effort to determine where, when (or how frequently), and how large (or how destructive) landslides are expected to be in an area. This information may prove useful to design landslide mitigation strategies, and to reduce landslide risk and societal and economic losses.
In the geomorphology literature, most of the attempts at predicting the occurrence of populations of landslides by adopting statistical approaches are based on the empirical observation that landslides occur as a result of multiple, interacting, conditioning and triggering factors. 
Based on this observation, and under the assumption that at the spatial and temporal scales of our investigation individual landslides are discrete ``point" events in the landscape, we propose a novel Bayesian modelling framework for the prediction of the spatio-temporal occurrence of landslides of the slide type caused by weather triggers. 
We build our modelling effort on a Log-Gaussian Cox Process (LGCP) by assuming that individual landslides in an area are the result of a point process described by an unknown intensity function. The modelling framework has two stochastic components: \i a Poisson component, which models the observed (random) landslide count in each terrain subdivision for a given landslide ``intensity'', \ie, the expected number of landslides per terrain subdivision (which may be transformed into a corresponding landslide ``susceptibility''); and \ii a Gaussian component, used to account for the spatial distribution of the local environmental conditions that influence landslide occurrence, and for the spatio-temporal distribution of ``unobserved" latent environmental controls on landslide occurrence. 
We tested our prediction framework in the Collazzone area, Umbria, Central Italy, for which a detailed multi-temporal landslide inventory covering the period from before 1941 to 2014 is available together with lithological and bedding data. We subdivided the $79~km^2$ area into $889$ slope units (SUs). In each SU, we computed the percentage of $16$ morphometric covariates derived from a $10$ $m$ $\times$ $10$ $m$ digital elevation model, and $13$ lithological and bedding attitude covariates obtained from a 1:10,000 scale geological map. We further counted how many of the 3,379 landslides in the multi-temporal inventory affect each SU and grouped them  into six periods. We used this complex space-time information to prepare five models of increasing complexity. Our ``baseline" model (Mod1) carries the spatial information only through the covariates mentioned above. It does not include any additional information about the spatial and temporal structure of the data, and it is therefore equivalent to a ``traditional" landslide susceptibility model. The second model (Mod2) is analogous, but it allows for time-interval-specific regression constants. Our next two models are more complex. In particular, our third model (Mod3) also accounts for latent spatial dependencies among neighboring SUs. These are inferred for each of the six time intervals, to explain variations in the landslide intensity and susceptibility not explained by the thematic covariates. By contrast, our fourth model (Mod4) accounts for the latent temporal dependence, separately for each SU, disregarding neighboring influences. Ultimately, our most advanced model (Mod5) contextually features all these relations. It contains the information carried by morphometric and thematic covariates, six time-interval-specific regression constants, and it also accounts for the latent temporal effects between consecutive slope instabilities at specific SUs as well as the latent spatial effects between adjacent SUs. This advanced model largely fulfills the definition of landslide hazard commonly accepted in the literature.
We quantified the spatial predictive performance of each of the five models using a $10$-fold cross-validation procedure, and the temporal predictive performance using a leave-one-out cross-validation procedure. We found that Mod5 performed better than the others. We then used it to test a novel strategy to classify the model results and to prepare a combined intensity--susceptibility landslide map, which provides more information than traditional susceptibility zonations for land planning and management. 
We discuss the advantages and limitations of the new modelling framework, and its potential application in other areas, making specific and general hazard and geomorphological considerations. We also give a perspective on possible developments in landslide prediction modelling and zoning.  
We expect our novel approach to the spatio-temporal prediction of landslides to enhance the currently limited ability to evaluate landslide hazard and its temporal and spatial variations. We further expect it to lead to better projections of future landslides, and to improve our collective understanding of the evolution of complex landscapes dominated by mass-wasting processes under multiple geophysical and weather triggers. 
 
\baselineskip=10pt

\par\vfill\noindent


\newpage

{\hypersetup{hidelinks}
\tableofcontents
}

\newpage

\baselineskip=16pt

\section{Introduction}
\label{sec:Introduction}

Landslides are ubiquitous in the hills, mountains, and high coasts that constellate the landmasses \citep{guzzetti2012}, and in many areas they cause significant human, societal, economic, and environmental damage and costs \citep{brabb1989,brabb1991,nadim-GlobalLandslide-2006,dowling2014, badoux2016,grahn2017,kirschbaum2009,pereira2017,froude2018,salvati-GenderAge-2018,rossi-PredictiveModel-2019}. The reliable anticipation of landslides and their consequences is thus of primary importance.

Like for other natural hazards, the anticipation of landslides involves predicting ``where" landslides can be expected (spatial prediction), ``when" or how frequently they can be expected (temporal prediction), and ``how many", how large or destructive one should expect the landslides to be in an area (number, size, impact, destructiveness prediction) \citep{varnes1984,guzzetti2005probabilistic,galli-LandslideVulnerability-2007,Tanyas2018}. The combined anticipation of ``where", ``when" (or how frequently), and ``how large" or destructive a landslide is expected to be, is called ``landslide hazard" \citep{cruden1997,hungr1999,guzzetti2005,guzzetti2005probabilistic,reichenbach2005,fell2008guidelines,lari2014}. Differently from other natural hazards, two distinct types of predictions are possible for landslides, namely, \i the prediction of single landslides, \ie, the anticipation of the behaviour of a single slope, or a portion of a slope, and \ii the prediction of populations of landslides, \ie, the anticipation of the behaviour of many (tens to several tens of thousands) landslides occurring in an area, and their spatial and temporal evolution. In this work, we focus on the prediction of populations of landslides in an area, and we do not consider the anticipation of the behaviour of single slopes or individual landslides. For this purpose, we exploit a unique multi-temporal inventory of landslides occurred over a multi-decade period in an area of Central Italy, which we use to fit and validate a set of five Bayesian geostatistical models constructed under the general assumption that landslides are a stochastic point process \citep{cox1965,cox1980,Chiu.al.2013}.

The paper is organised as follows. We begin, in  \S\ref{sec:landslide-prediction}, by providing background information on traditional spatial and temporal landslide predictive modelling approaches, and their limitations. This is followed, in  \S\ref{StudyArea}, by a description of the study area of Collazzone, Italy, and, in  \S\ref{sec:data}, of the landslide, the morphological, and the thematic data used, of our choice of the modelling mapping unit, and the pre-processing steps. Next, in  \S\ref{sec:LGCP}, we describe five different geostatistical models that we have implemented, consisting of: \i a baseline model where the landslide spatial dimension is only carried through the explanatory variables; \ii an improved version of the baseline model which allows for time-interval-specific regression constants; and three extensions to the second baseline model which account for \iii spatial, \iv temporal, and \v spatio-temporal random effects acting at a latent level. This is followed, in  \S\ref{sec:results}, by the presentation and comparison of the results for the five geostatistical models, and the associated calibration and validation diagnostics. In \S\ref{sec:Discussion}, we discuss the results obtained, and we provide geomorphological insight on the performed statistical inference. Lastly, in \S\ref{sec:Conclusions}, we summarise the lessons learnt and we outline the remaining challenges.

\section{Prediction of landslide occurrence}
\label{sec:landslide-prediction}

In the geomorphological literature, most of the attempts at predicting the occurrence of populations of landslides in an area are based on the empirical observation that landslides are spatially and temporally discrete events that occur as a result of multiple, interacting, conditioning and triggering factors. The conditioning factors primarily influence where landslides can occur, whereas the triggering factors drive the landslides onset, \ie, the time or period of occurrence of the slope failures. Together, the conditioning and the triggering factors control the extent of the area affected by landslides and the size distribution of the slope failures, which is linked to the landslide impact and destructiveness. Because of the complexity and the variability of the landslide processes, which depend among others on the soil, rock, and other landscape characteristics, and on the weather or seismic triggers, and because the exact or even approximate locations of the landslides are unknown before they occur, individual slope failures in a population of landslides can be considered as the realisation of a stochastic process \citep{Das2012,lombardo2014test}, and modelled accordingly. 

A large variety of approaches have been proposed to assess the landslide ``susceptibility'', which refers in the geomorphological literature to the spatially-varying, time-independent likelihood of landslides occurring in an area given the local terrain conditions \citep{brabb1985,chung-ProbabilisticPrediction-1999,guzzetti1999landslidehazard,guzzetti2005probabilistic,reichenbach2018}. These approaches can be loosely grouped into five main categories \citep{guzzetti2005}, \ie, \i direct geomorphological mapping \citep{verstappen1983,hansen1995,reichenbach2005}, \ii analysis of landslide inventories \citep{campbell1973,degraff1988,moreiras2004}, \iii heuristic, index-based methods \citep{nilsen1977,posner2015}, \iv deterministic, physically-based, conceptual models \citep{ward1981,ward1982,montgomery1994,dietrich2001validation,Bout2018}, and \v statistical prediction models \citep{carrara1983,chung-ProbabilisticPrediction-1999,guzzetti1999landslidehazard,van2006,lombardo2016a,reichenbach2018}. Each of these approaches has potential advantages and inherent limitations \citep{guzzetti2005,van2006,lombardo2015binary}. 

Geomorphological mapping depends entirely on the skills and experience of the investigators. It may provide reliable results, but it is difficult to reproduce, impractical over large areas, and inadequate for quantitative hazard assessments \citep{guzzetti1999landslidehazard,van1999}. Analysis of the inventories depends on the quality and completeness of the available landslide maps \citep{TanyasLombardo2019}. Where an inventory is incomplete, or wrong, the susceptibility assessment will be underestimated, or biased \citep{guzzetti2012}. Heuristic methods rely on the (often unproven) assumption that all the causes for landslides in an area are known, and they produce qualitative and subjective predictions unsuited for quantitative hazard assessments \citep{soeters1996,leoni2009gis}. Physically-based models exploit the existing understanding of the mechanical laws that control slope instability. Their limitation lies in the inherent simplicity of the modelling equations that may not capture the complex interactions controlling the slope stability/instability conditions. Furthermore, the physically-based models require large datasets to describe the surface and subsurface mechanical and hydrological properties of the terrain, which are difficult and expensive to obtain. As a result, physically-based models are used chiefly for small or very small areas \citep[\eg,][]{montgomery1994,chakraborty2016,seyed2019}, albeit a few examples also exist of applications for large areas \citep[\eg,][]{gorsevski2006,Raia2014,alvioli2016parallelization}.

Lastly, statistical approaches aim at exploiting the ``functional" relationships existing between a set of instability factors, and the past and present distribution of landslides obtained typically from a landslide inventory map \citep{carrara1983,duman2005,guzzetti2012}, or a landslide catalogue \citep{van2012}. The large number of statistically-based approaches proposed in the literature \citep{reichenbach2018} almost invariably exploit classification methods, and provide probabilistic estimates suited for quantitative hazard assessments. Statistical models can be constructed using a variety of thematic and environmental variables obtained from existing maps or by processing remotely sensed images and data, in different landscape and environmental settings, covering a broad range of scales and study-area sizes. The dependent variable is obtained from different types of landslide inventory maps \citep{guzzetti2012} or landslide catalogues \citep{van2012}, and is typically used in a binary structure, expressing the presence or absence of landslides in each mapping unit, where a terrain mapping unit is a regular or irregular geographical subdivision (\eg, a pixel, unique condition, slope or hydrological unit, administrative subdivision \citep{guzzetti1999landslidehazard,guzzetti2005,van2006}) used to partition a study area. The fitted model is then used to assess the landslide susceptibility for each mapping unit
\citep{guzzetti2005}.

Focusing on statistically-based susceptibility approaches, a limitation of the traditional and of most of the current models is that they predict only whether a mapping unit is expected to have (or not have) landslides, regardless of the number or size of the expected failures in each unit. In a population of landslides, the size (\ie, length, width, area, volume) of the slope failures is linked to the number of the failures. \citet{hovius1997} and \citet{malamud2004}, among others, have shown that the area of a landslide obeys empirical probability distributions that control the average size and relative proportion of the landslides in an area. Similar empirical dependencies have been found for landslide volume \citep{brunetti2009}, for the landslide area to volume ratio \citep{guzzetti2009,larsen2010}, and more recently for the landslide width to length ratio \citep{taylor2018}. 

In an attempt to overcome the inherent inability of landslide susceptibility models to predict the number of the expected landslides, \citet{Lombardo.etal:2018} have proposed a novel framework for landslide intensity assessment.  The approach treats single landslides in a population of landslides as individual realisations from a continuous--space point process \citep{cox1980,Chiu.al.2013}. This is different from the discrete-space binary presence-absence model adopted by the traditional statistically-based susceptibility models \citep{carrara1983,guzzetti1999landslidehazard,guzzetti2005,lombardo2018presenting,reichenbach2018}. As a result, the approach aims at predicting the number of the expected landslides in each mapping unit adopting a Poisson distribution, whose mean is linked to the unknown landslide intensity that can be estimated from a landslide dataset. The approach was applied successfully to model populations of rainfall--induced  \citep{Lombardo.etal:2018,lombardo2019numerical} and seismically--triggered \citep{lombardo2019geostatistical} landslides in different morphological, geological, and climatic settings. 

A second limitation of most statistically-based models is that they typically do not consider the spatial relationships between landslide occurrences in different terrain mapping units. Landslide occurrences are often assumed to be independent given the terrain conditions. In other words, the geographical location of the mapping units is not explicitly taken into account, so that adjacent, neighbouring, and distant units are considered equally by the models. Not considering the spatial dependencies (or lack thereof) among units placed in different locations in a study area may result in poorer susceptibility models and associated zonations \citep{reichenbach2018,Lombardo.etal:2018}. 

A third limitation of most statistically-based models is the fact that they consider the likelihood of landslides occurring in an area to be constant in (\ie, independent of) time \citep[\eg,][]{meusburger2009,cama2015predicting}. In the long run (\ie, hundreds to thousands of years), the assumption does not hold because the landslide triggering conditions (\eg, the frequency of high or prolonged rainfall periods, of snow melt events, or of earthquakes), as well as the predisposing conditions (\eg, land use or land cover) may change with time, inevitably changing the landslide susceptibility. Recent empirical evidence shows that in some landscapes, even for short periods (\ie, tens of years), when a landslide occurs it may become an ``attractor" for future landslides, with new landslides being more likely to occur inside or in the immediate vicinity of the previous landslides. \citet{samia-CharacterizationQuantification-2017,samia-LandslidesFollow-2017} called this effect  ``landslide path dependency'', and found that the spatio-temporal dependency of new landslides on old landslides disappears in their study area, in Umbria, Central Italy---the same study area of this work---approximately after 10--15 years. The evidence violates the assumption that susceptibility is constant in time, even for short periods. 

In the literature, approaches to predict the temporal occurrence of landslides are equally if not more diversified. Depending on the scope, the temporal coverage, the return time of the predictions, and the extent of the study area, methods include \i the use of empirical rainfall thresholds for the possible occurrence of landslides \citep{glade2000,dai2001,crosta-RainfallThresholds-2000,aleotti2004warning,guzzetti2007,guzzetti2008,saito2010,ko2018,segoni-ReviewRecent-2018,guzzetti-LEWS-2019}, \ii physically-based, coupled, distributed rainfall and infiltration slope stability models \citep{montgomery1994,vanH1996,baum-TRIGRSFortran-2008,lanni2013,formetta2014,reid-Scoops3DSoftware-2015,Formetta2016,alvioli2016parallelization,Bout2018}, and \iii the analysis of time series of historical landslides and landslide events \citep{crovelli2009,rossi2010,witt2010}. Only physically-based models (inherently) consider the spatio-temporal interactions that condition landslide occurrence, which are not considered by the threshold models or by the analysis of the historical records. However, the physically-based models are generally applicable only to small areas and for short periods of time (a few hours to a few days), and are not suited for the spatio-temporal modelling of landslide occurrence over large areas and for long periods;  is essential for medium to long term land planning and management.

In this work, we construct innovative geostatistical models that consider \i spatial,  \ii temporal, and \iii spatio-temporal landslide latent effects among: adjacent terrain mapping units, same mapping units but subsequent in time and both conditions together, respectively. We consider this an improvement over the existing approaches to predict the spatio-temporal occurrence of landslides in an area.

\section{Study area}
\label{StudyArea}

The area of Collazzone, Umbria, Central Italy, is well studied and represents a unique site where landslides have been mapped repeatedly over a large timespan. A description of the area can be found in \citet{Guzzetti2006a,Guzzetti2006b}, \citet{Ardizzone2007}, \citet{galli-ComparingLandslide-2008} and other references therein. Overall, our study area extends over $\approx$~79 $km^2$, with terrain elevation between 145 $m$ along the Tiber River flood plain NNW of Todi, and 634 $m$ at Monte di Grutti (Figure~\ref{fig:StudyArea}). The landscape is predominantly hilly, and lithology and the attitude of bedding planes control the morphology of the slopes, and the presence and abundance of the landslides. In the area  sedimentary rocks crop out, Cretaceous to Recent in age, including recent fluvial deposits, continental gravel, sand and clay, travertine, layered sandstone and marl, and thinly layered limestone. The climate is Mediterranean, with precipitation falling mostly in the period from October to December, and from February to May. Intense rainfall events or prolonged rainfall periods are the primary natural triggers of landslides in the area \citep{ardizzone2013very}, followed by the rapid melting of snow \citep{cardinali2000landslides}. Landslides are abundant and cover 17.05 $km^2$---corresponding to a density of $\approx$ 43 landslides per square kilometre---and range in age, type, morphology, and volume from very old, partly eroded, large and deep-seated slides and slide-earth flows, to young and shallow slides and flows \citep{Hungr2014}. Recent landslides are most abundant in the cultivated areas and are rare in the forested terrain, indicating a dependence of the recent landslides on the agricultural practices \citep{mergili2014spatially}.

\section{Data compilation and pre-processing}
\label{sec:data}

\subsection{Landslide data}
\label{Landslides}

We obtained the landslide information from a pre-existing multi-temporal landslide inventory prepared at 1:10,000 scale through the systematic visual interpretation of five sets of aerial photographs taken in 1941, 1954, 1977, 1985, and 1997, and of five stereoscopic, panchromatic and multi-spectral satellite image pairs acquired in August 2009, March 2010, May 2010, April 2013, and April 2014 (Table~\ref{tab:images}), supplemented by field checks and surveys executed in various periods from 1998 to 2004, in May 2004, and in December 2005 \citep{Guzzetti2006a,Ardizzone2007,galli-ComparingLandslide-2008,ardizzone2013very}. The multi-temporal inventory shows the location, surface area, type  \citep{Hungr2014}, and estimated age \citep{cruden-LandslideTypes-1996} of 3,379 landslides, ranging in size from $A_L$ = $5.8\times10^3$ $m^2$ to $A_L$ = $1.5\times10^6$ $m^2$ (mean, $\mu=6.9\times10^3$ $m^2$, standard deviation, $\sigma=3.2\times10^4$ $m^2$), for a total area covered by landslides of $A_{\text{LT}}$ $=$ 17.05 $km^2$. In the inventory, landslide age was defined as very old (relict), old (predating 1941), or recent (from 1941 to 2014), using photo-interpretation criteria and field evidence, despite some ambiguity in the definition of the age of a landslide based on its morphological appearance \citep{mccalpin1984,Guzzetti2006a}.

\begin{table}[!htbp]
\centering
\caption{Main characteristics of the aerial photographs and the optical satellite images used to prepare the multi-temporal landslide inventory for the Collazzone study area, Umbria, Central Italy, used in this work and shown in Figure~\ref{fig:StudyArea}. Legend: GSD, ground sampling distance.}
\label{tab:images}
\begin{tabular}{llllll}
& &  &  & &    \\ \hline
Year & Period  & Type  & Nominal scale & Mode & Source \\ \hline
1941 & Summer      & Panchromatic & 1:18,000   &  Stereo & Aerial photographs \\
1954 & Spring-Summer & Panchromatic & 1:33,000   &  Stereo & Aerial photographs   \\
1977 & Spring-Summer      & Colour       & 1:13,000   &  Stereo & Aerial photographs   \\
1985 & July             & Panchromatic & 1:15,000    &  Stereo & Aerial photographs  \\
1997 & April            & Panchromatic & 1:20,000   &  Stereo & Aerial photographs   \\
2009 & 12 August    & Panchromatic & GSD=0.41 $m$  &  Stereo  & GeoEye-1   \\
2010 & March          & Panchromatic & GSD=0.50 $m$      &  Stereo  & WorldView-1  \\
2010 & 27 May        & Panchromatic & GSD=0.41 $m$    &  Stereo  & GeoEye-1\\
2013 & 13 April        & Panchromatic & GSD=0.50 $m$   &  Stereo   & GeoEye-1\\
2014 & 14 April        & Multispectral   & GSD=2.00 $m$   &  Stereo & WorldView-2  \\ \hline
\end{tabular}
\end{table}

\begin{figure}[!htbp]
	\centering
	\includegraphics[width=0.73\textwidth]{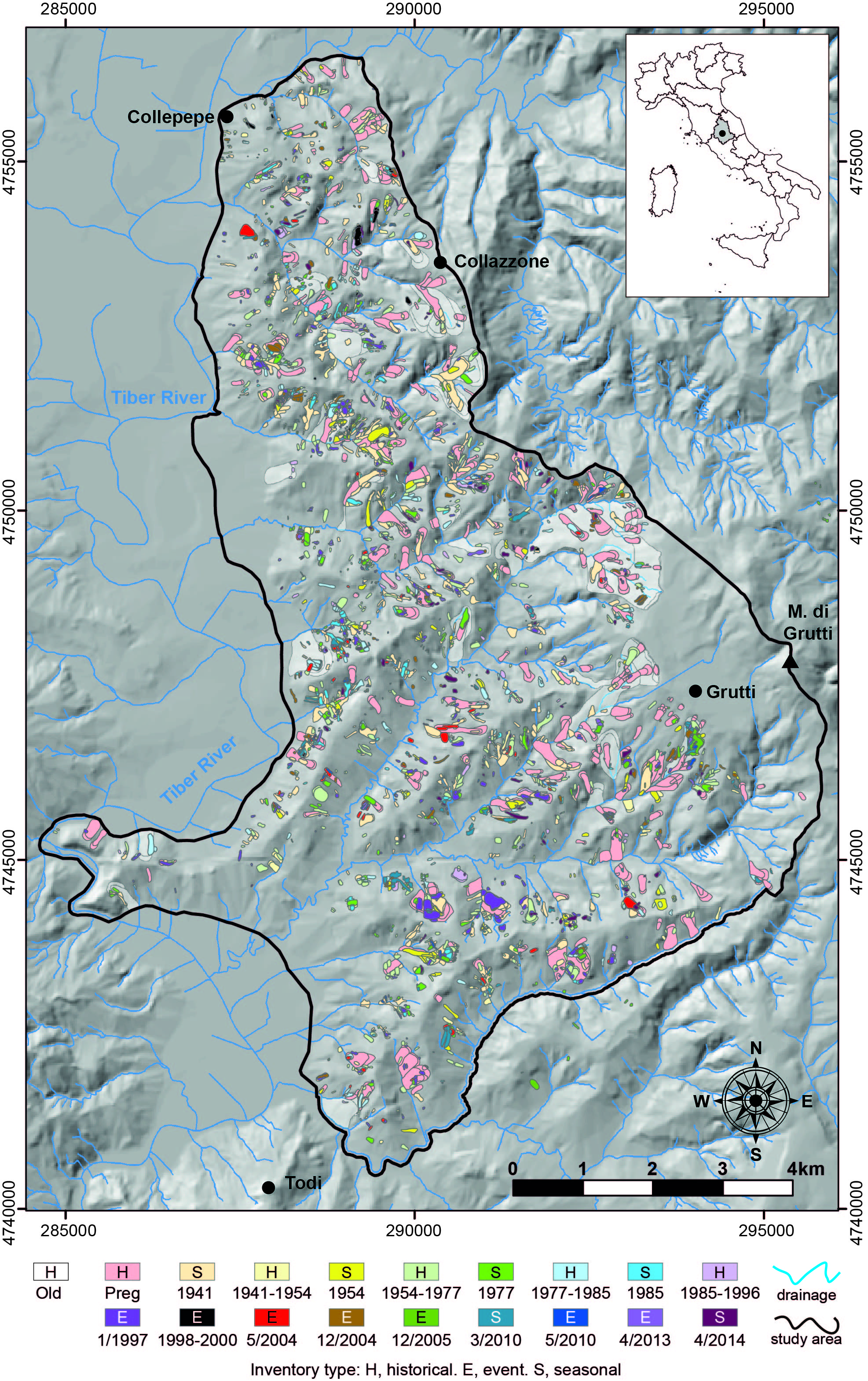}
	\caption{Collazone study area, Umbria, Central Italy.  The map shows the multi-temporal landslide inventory, morphology, and hydrology of the study area. Coloured polygons are landslides of various ages or periods mapped through the visual interpretation of aerial photographs or satellite images of different vintages, or through field work. Individual inventories shown by different colours are of three types: E, event; S, seasonal; H, historical. See text for further explanation.}
\label{fig:StudyArea}
\end{figure}

Key to our study is the fact that in the multi-temporal inventory landslides are separated into nineteen, irregularly spaced ``temporal slices'', each shown by a different colour in Figure~\ref{fig:StudyArea}. Landslides in the temporal slices before 2000 were detected and mapped using black and white (panchromatic) and colour (1977) aerial photographs, whereas landslides between 2009 and 2014 were detected and mapped using VHR stereoscopic satellite images (Table~\ref{tab:images}), and directly in the field. Visual interpretation of the stereoscopic satellite images was performed using different software, including: \i ERDAS IMAGINE\textsuperscript{\textregistered} and Leica Photogrammetry Suite (LPS) for block orientation of the stereo images; \ii Stereo Analyst for ArcGIS\textsuperscript{\textregistered} for image visualization and landslide mapping; and \iii StereoMirror\textsuperscript{\texttrademark} hardware technology to obtain 3D views of the VHR satellite images. 

The same interpretation criteria were adopted to identify and map the landslides on aerial photographs and the satellite images. \citep{guzzetti2012,ardizzone2013very,murillo2015}. In each set of aerial photographs and in each pair of satellite images, landslides that appeared ``fresh'' were given the date (\ie, year) of the aerial photographs, or the date (\ie, month and year) of the satellite images. The ``non-fresh'' landslides were attributed to the period (\ie, the ``temporal slice") between two successive sets of aerial photographs or satellite image pairs. Landslides mapped in the field after single or multiple rainfall events between 1998 and 2013 were given the date (\ie, month and year) of the field surveys. The different methods and instruments used to interpret the aerial photographs and the satellite images, and the fact that some of the landslides were mapped in the field, have conditioned the completeness and accuracy of the landslide information in the multi-temporal inventory, which is therefore not constant throughout the time slices. In general, more recent time slices showing event or seasonal inventories (E and S, respectively, in Figure~\ref{fig:StudyArea}) have more numerous landslides of small sizes, whereas historical inventories (H, in Figure~\ref{fig:StudyArea}) show larger landslides, and lack landslides of very small size. This is a known bias of the multi-temporal inventory used in our study \citep{Guzzetti2006a,Ardizzone2007,galli-ComparingLandslide-2008,ardizzone2013very}. 

\subsection{Mapping unit}
\label{sec:Units}

\begin{figure}[!htbp]
	\centering
	\includegraphics[width=\linewidth]{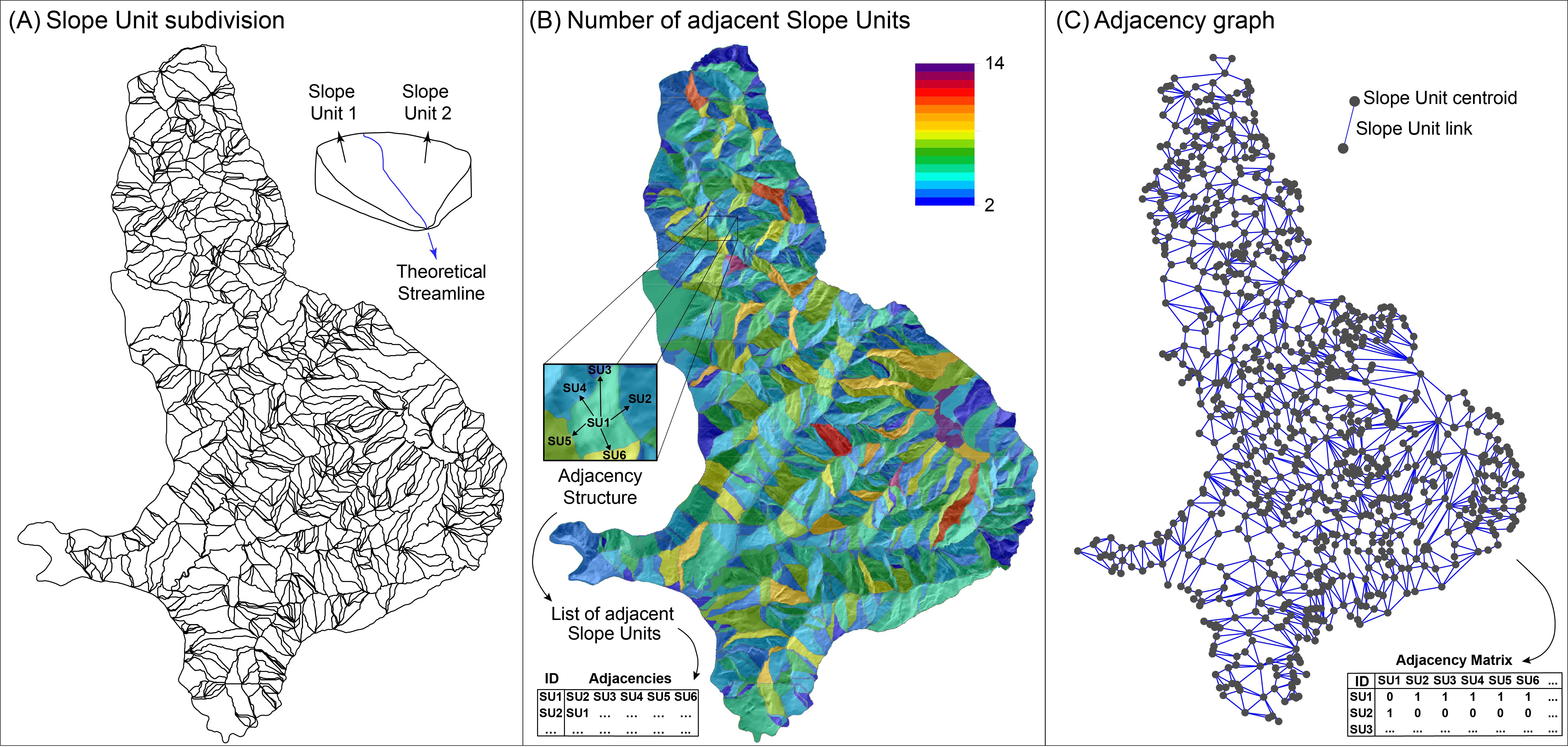}
	\caption{Collazzone study area, Umbria, Central Italy. 
	(A) Geographical subdivision of the study area into 889 slope units (SUs).
	(B) Map showing number of adjacent SUs, and adjacency structure.
	(C) Adjacency graph showing links (blue lines) between adjacent SU centroids (grey dots), and adjacency matrix.}
	\label{fig:AdjMat}
\end{figure}

Prediction of landslide occurrence in an area requires the preliminary selection of a suitable terrain mapping unit, \ie, a subdivision of the terrain that maximises the within-unit (internal) homogeneity and the between-unit (external) heterogeneity across distinct physical or geographical boundaries 
\citep{Carrara1991,guzzetti2005}. In agreement with previous studies in the same  \citep{Guzzetti2006b}, in similar \citep{carrara2003}, and in different \citep{VanDenEeckhaut2009,Erener2012,Camilo2017,amato2019accounting} areas, we select the ``slope unit'' (SU) as the mapping unit of reference. By definition, a SU is a terrain geomorphological unit bounded by drainage and divide lines, and corresponds to a slope, a combination of adjacent slopes, or a small catchment  \citep{Carrara1991,alvioli2016automatic}. We use a subdivision of the study area (\ie, our spatial domain) into 889 SUs ranging in size from $A_L$ = $6.17\times10^2$ $m^2$ to $A_L$ = $1.4\times10^6$ $m^2$ (mean, $\mu=8.9\times10^4$ $m^2$, standard deviation, $\sigma=1.1\times10^5$ $m^2$), corresponding to an average density of one SU approximately every $0.1$ $km^2$. Panel A of Figure~\ref{fig:AdjMat} portrays the geographical distribution of the SUs used in the study.

\subsection{Thematic data and explanatory variables}
\label{sec:covariates}

To support the modelling procedure, we used a set of morphometric and thematic variables (also called ``covariates"), which we list in Table~\ref{tab:covariates}. The 16 morphometric covariates were derived from a 10 $m$ $\times$ 10 $m$ resolution digital elevation model (DEM) obtained through the automatic interpolation of 10 $m$ and 5 $m$ interval contour lines taken from 1:10,000 scale topographic base maps made accessible by the Umbria Region government (\href{http://www.umbriageo.regione.umbria.it/pagina/distribuzione-carta-tecnica-regionale-vettoriale-1}{link here}). To represent morphometric covariates at the SU scale, we  computed the two main statistical properties (mean $\mu$, and standard deviation $\sigma$) of each covariate for each respective SU. We further obtained the thematic covariates (Table~\ref{tab:covariates}) from a geological map prepared at 1:10,000 scale by \citet{Guzzetti2006a}, and used in previous landslide susceptibility and hazard studies in the same area \citep{Guzzetti2006a,Ardizzone2007,galli-ComparingLandslide-2008}. From the geological map we extracted information on nine lithological units, and four bedding domain classes. We selected these covariates because bedrock geology and the attitude of bedding planes have been shown to control the presence (or absence) and the abundance of landslides in our study area \citep{Guzzetti2006a,marchesini2015assessing}. To summarise the geologic and bedding information for each SU, we computed the ratio between the relative extent of each categorical class in a given SU and the total extent of the SU. As a result, we transformed the original categorical information into a continuous one, expressing the proportion of area coverage per class in each SU.

\begin{table}[!htbp]
\centering
\caption{Morphometric (M), lithological (L), and bedding-related (structural, S) explanatory variables (covariates) used in the study for space-time landslide predictive modelling in the Collazzone area, Umbria, Central Italy (see Figure~\ref{fig:StudyArea}). References: 
1, \href{http://www.umbriageo.regione.umbria.it/pagina/distribuzione-carta-tecnica-regionale-vettoriale-1}{http://www.umbriageo.regione.umbria.it/pagina/distribuzione-carta-tecnica-regionale-vettoriale-1};
2, \citet{zevenbergen1987quantitative};
3, \citet{Lombardo2018};
4, \citet{heerdegen1982quantifying};
5, \citet{bohner2006spatial};
6, \citet{beven1979physically};
7, \citet{Guzzetti2006a}.}
\label{tab:covariates}
\begin{tabular}{lllll}
& & & &    \\ \hline
Type & Variable & Description  & Source & Reference \\ \hline
\vspace{-0.35cm}
& & & &      \\
M & ELEV$\mu$ & Terrain elevation mean & 10 $m$ $\times$ 10 $m$  DEM  & 1  \\
M & ELEV$\sigma$ & Terrain elevation st.dev. & 10 $m$ $\times$ 10 $m$  DEM  & 1 \\
M & SLOPE$\mu$ & Terrain slope mean & 10 $m$ $\times$ 10 $m$  DEM &  2  \\
M & SLOPE$\sigma$ & Terrain slope st.dev. & 10 $m$ $\times$ 10 $m$  DEM &     2    \\
M & ENES$\mu$ & Eastness mean & 10 $m$ $\times$ 10 $m$  DEM  & 3   \\
M & ENES$\sigma$ & Eastness st.dev. & 10 $m$ $\times$ 10 $m$  DEM  & 3    \\
M & NNES$\mu$ & Northness mean & 10 $m$ $\times$ 10 $m$  DEM  & 3    \\
M & NNES$\sigma$ & Northness st.dev. & 10 $m$ $\times$ 10 $m$  DEM  & 3   \\
M & PLCR$\mu$ & Planar curvature mean & 10 $m$ $\times$ 10 $m$  DEM  & 4   \\ 
M & PLCR$\sigma$ & Planar curvature st.dev. & 10 $m$ $\times$ 10 $m$  DEM  & 4 \\
M & PRCR$\mu$ & Profile curvature mean & 10 $m$ $\times$ 10 $m$  DEM  & 4 \\
M & PRCR$\sigma$ & Profile curvature st.dev. & 10 $m$ $\times$ 10 $m$  DEM & 4\\
M & RSP$\mu$ & Relative slope position mean & 10 $m$ $\times$ 10 $m$  DEM  & 5 \\
M & RSP$\sigma$ & Relative slope position st.dev. & 10 $m$ $\times$ 10 $m$ DEM & 5 \\
M & TWI$\mu$ & Topographic wetness index mean & 10 $m$ $\times$ 10 $m$  DEM & 6 \\
M & TWI$\sigma$ & Topographic wetness index st.dev. & 10 $m$ $\times$ 10 $m$ DEM & 6 \\
\vspace{-0.35cm}
& &  & &   \\
L & AD\_R  & Alluvial sediment & Lithological map, 1:10,000 & 7 \\
L & C\_R  & Clay & Lithological map, 1:10,000  & 7 \\
L & G\_R  & Gravel & Lithological map, 1:10,000 & 7 \\
L & L\_R  & Limestone & Lithological map, 1:10,000 & 7 \\
L & M\_R  & Marl & Lithological map, 1:10,000  & 7 \\
L & S\_R  & Sand & Lithological map, 1:10,000  & 7 \\
L & SGC\_R  & Gravel, sand, clay & Lithological map, 1:10,000 & 7  \\
L & SS\_R  & Sandstone & Lithological map, 1:10,000  & 7  \\
L & T\_R  & Travertine & Lithological map, 1:10,000  & 7  \\
\vspace{-0.35cm}
& & & &   \\
S & AS\_R  & Anaclinal slope & Bedding map, 1:10,000   & 7 \\
S & CS\_R  & Cataclinal slope & Bedding map, 1:10,000  & 7  \\
S & OS\_R  & Orthogonal slope & Bedding map, 1:10,000 & 7   \\
S & US\_R  & Unbedded sediment & Bedding map, 1:10,000 & 7  \\ \hline
\end{tabular}
\end{table}

\subsection{Pre-processing}
\label{sec:pre-processing}

\begin{figure}[!htbp]
\centering
	\includegraphics[width=\linewidth]{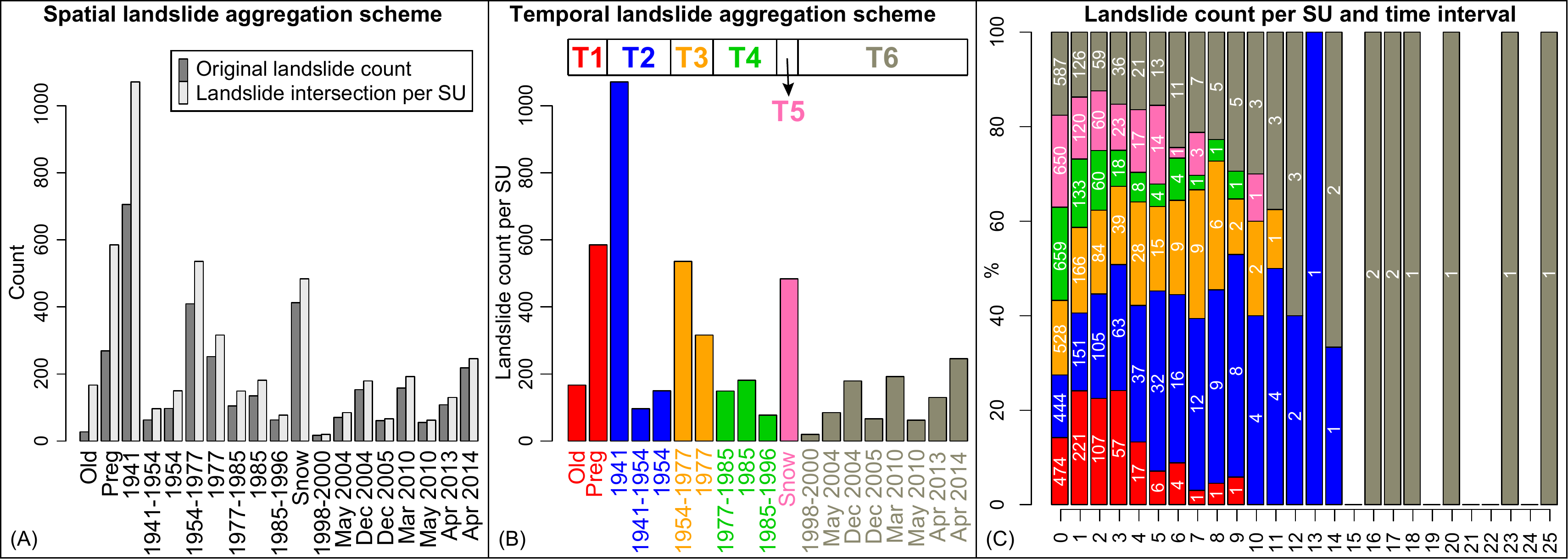}
	\caption{Collazzone study area, Umbria, Central Italy.
    Spatial and temporal aggregation schemes of landslide counts.
	(A): Temporal distribution of the original counts obtained by considering landslides as pure points (dark grey); and, spatial aggregation by considering the areal extent of landslides (light grey). Hence, where landslides covered more than one SU, we repeated the count over multiple SUs to respect the geomorphological and areal realization of the instability process. 
	(B): Spatially aggregated landslide count per SU displayed per ``temporal slices" for the 19 inventories shown in Figure~\ref{fig:StudyArea}; 
	and, associated temporal aggregation scheme to merge the 19 time slices (single bars) into six ``time periods" (groups of bars shown by the same colour). 
	(C): Summary staked bar chart of the aggregated landslide counts distribution per SU for six near-regular time periods, from T1 to T6: the x-axis reports the SUs' rank, from SUs having zero landslides (left) to SUs having up to 25 landslides (right). The y-axis shows the proportion of landslide counts across the six different time periods, shown by different colours. White characters correspond to the counts per time interval.
	More information on SUs is provided in Section \ref{sec:Units}.}
	\label{fig:AggrByTime}
\end{figure}

We initially tested our modelling framework (see Section~\ref{sec:LGCP}) using the 19 separate time slices shown in Figure~\ref{fig:StudyArea}, where the original 19 landslide count distributions represented our target variable. However, these preliminary tests did not produce satisfying predictive performances (unreported results). Possible reasons for the unsatisfactory results included \i the irregular time-span of the landslide inventories, ranging from 1 to 23 years, \ii the highly variable number of landslides in each time slice, from 303 landslides for the 1985-1996 time slice, to 866 landslides for the 1941-1954 time slice, and \iii the fact that some time slices have only a few landslides (Figure~\ref{fig:AggrByTime}A, dark grey). To address the issue, we aggregated the landslides both in space and time.

To aggregate the data over space, we looked into the inventories and realized that some of the landslides have a large areal extent at certain times, leading the instability to affect several SUs simultaneously. In such cases, treating these landslides as a single points (\ie, assigning one count to a single SU) would neglect the areal extent of the landslide phenomenon. Therefore, in cases where a landslide covers more than one SU, we assigned a count to multiple SUs under the constraint that the intersections between the landslide and the SUs are larger than 2\%  of the area of the SU \citep{carrara2013}. In turn, this procedure generated a moderately larger landslide count per SU than the original landslide count, as shown in Figure~\ref{fig:AggrByTime}A. 

The second aggregation scheme involved the temporal dimension of the spatially-aggregated landslide counts. Specifically, we further aggregated the 19 temporal slices in a nearly regular temporal grid with intervals of approximately 15 years, with the exception of the stand-alone snow-melt event. The procedure produced six ``time intervals", each composed of one to eight time slices, which we will refer to as T1 to T6 in the rest of the manuscript.  Panel B of  Figure~\ref{fig:AggrByTime} illustrates the distribution of landslide counts per SU for each time interval, and Panel C of  Figure~\ref{fig:AggrByTime} summarises our final aggregation scheme.

In addition to the covariate preparation and the spatio-temporal aggregation, the pre-processing phase involved creating the so-called adjacency matrix \citep{zhang2010shortest}. The $(i_1,i_2)$-th entry of the adjacency matrix is equal to one if the $i_1$-th SU shares a border with the $i_2$-th SU, and it is zero otherwise. Therefore, it is a symmetric matrix (\ie, if SU1 is adjacent to SU2, then SU2 is also adjacent to SU1), which indicates the neighbourhood structure between SUs, and it provides fundamental information to build the space and space-time models presented in this work. Maps (B) and (C) in Figure~\ref{fig:AdjMat}  summarise the main steps involved in the calculation of the adjacency matrix, and illustrate the adjacency graph structure.

\section{Modelling framework}
\label{sec:LGCP}

\subsection{Fundamentals of point processes}

Conceptually, we identify a landslide with a single point $(s_i,t_i)$, where $s_i$  is the spatial location and $t_i$ is time. In practice, we may also consider the spatial dimension of large landslides by counting them more than once. The time resolution here corresponds to the six time intervals, such that $t_i\in\{1,2,\ldots,6\}$. For the purpose of estimation and mapping at the resolution of the SUs, we do not need the exact position $s_i$ of landslides, but only the SU.

Our fundamental modelling assumption is that the space-time points $(s_i,t_i)$ identifying landslide occurrences stem from an unobserved intensity function $\lambda(s,t)$ that varies over space and time. The interpretation of this intensity function is that we observe (approximately) $\lambda(s,t)$ events on average per spatio-temporal unit around $(s,t)$.  For an arbitrary domain $D$ stretching over space and time, the mean number of observed events is given by the integral $\int_D \lambda(s,t) {\rm d}s{\rm d}t$. The natural distribution of such event counts is the Poisson distribution with intensity given by this integral. We assume that this stochastic mechanism generates the observed spatio-temporal point pattern, and we call it a \emph{Poisson point process}.

In practice, we construct a model for the intensity function $\lambda(s,t)$ which incorporates covariate effects, and  which may further capture structured variations of the space-time intensity surface of random nature, \ie, not explained by the available covariate information. Such random effects can be considered as ``unobserved" effects, hence covariates during the modelling process, whose signals influence the landslide space-time pattern in the data. For instance, in case of event-based landslide studies, the precipitation or seismic triggers may not be included among the covariate set (\eg, if they are unobserved), although they influence the concentration of landslides at specific locations. Nevertheless, advanced spatial models can retrieve the pattern of the unobserved trigger from the landslide distribution itself, as demonstrated by \citet{Lombardo.etal:2018,lombardo2019numerical} for storm-induced and by \citet{lombardo2019geostatistical} for earthquake-induced landslides. As for temporal unobserved effects, advanced temporal models can retrieve the influence of the ``landslide path dependency" recognised by \citet{Samia2018}. 

When allowing for such random components in the intensity function $\lambda$, the resulting stochastic model for the observed point pattern is called a \emph{Cox point process} \citep{cox1965}, and we use the uppercase notation $\Lambda(s,t)$ for the intensity function to highlight its stochastic behaviour. More specifically, if we opt for the flexible and convenient choice of additive random effects in the log-intensity $\log(\Lambda(s,t))$ possessing Gaussian process distributions, we obtain a \emph{Log-Gaussian Cox Process} \citep{Moller.al.1998,basu2002,diggle2013}, \emph{LGCP} in short. 

For our dataset, we can rewrite the general structure of such models by taking into account the spatial ($889$ SUs) and temporal ($6$ intervals) resolution and by using the surface area $|A_i|$, $i=1,\ldots,889$, of SUs. We make the natural assumption that the average number of landslides observed within a SU $s_i$ is proportional to its surface area $|A_i|$, given that all other predictor components are the same. Furthermore, we model a separate spatial intensity for each of the temporal intervals $j=1,\ldots,6$, and so we write $\Lambda_j(s_i)=|A_i|\times\tilde{\Lambda}_j(s_i)$ for the integrated intensity of the $i$-th SU in the $j$-th temporal interval, where $\tilde{\Lambda}_j(s_i)$ can be interpreted as the intensity of a (rescaled) unit-area SU with the same characteristics.

We further write $N_{ij}\in\{0,1,2,\ldots\}$ for the number of landslides observed in SU $i$ during temporal interval $j$. Then, given the intensity values $\Lambda_j(s)$, the model has the following structure:
\begin{equation}
N_{ij}\mid \Lambda_j(s) \sim \mathrm{Poisson}(|A_i|\times\tilde{\Lambda}_j(s_i)), \quad i=1,\ldots, 889, \quad j=1,\ldots, 6.
\end{equation}
We will formulate five regression models (technically speaking, Poisson regressions with a log-link function) in \S\ref{sec:Baseline}--\S\ref{sec:space-time}, with two variants of a baseline model without random effects presented in \S\ref{sec:Baseline}. These models integrate observed covariate effects and random effects into the log-intensity $\log(\Lambda_j(s))$ in an additive way. The models follow the general structure
\begin{equation}
\log(\Lambda_j(s_i))=\log(|A_i|)+\text{fixed covariate effects}+\text{random effects}
\end{equation}
where the random effect component are not accounted for in our baseline models. The term $\log(|A_i|)$ represents a fixed deterministic offset. 

\subsection{The Bayesian modelling paradigm}
\label{sec:BayesianModelling}

Before presenting the specific structure of the five models we tested, we first recall the general idea of fully Bayesian modelling as implemented here. We aim to simultaneously estimate the latent intensity function $\Lambda_j(s)$, and more specifically its components such as the coefficients of the fixed covariate effects and random effects, if present. Moreover, a relatively small number of so-called hyperparameters governing the smoothness and variance of the random effects is also calibrated automatically. In Bayesian modelling, we specify so-called prior probability distributions for the components and parameters to estimate. In general, prior distributions allow incorporating expert knowledge and can help stabilise estimated models when these have a very complex structure and/or when data are very noisy. Bayes' famous Theorem then tells us how we can construct the posterior distributions (densities, expected values, etc.), \ie, parameter estimates and precise probabilistic uncertainty statements, by confronting prior information with observed data. Therefore, the Bayesian mechanism allows us to move from the specification of ``data distribution given the model parameters and their prior information'' to the estimation target corresponding to the ``distribution of model parameters given the data and prior information''. The fitted posterior distributions can then be exploited to make inference (\eg, on model parameters and their uncertainty), derive any model-based diagnostics, and draw practical conclusions.

Through the specification of priors, the Bayesian paradigm allows us to resolve potential parameter identifiability issues by naturally integrating constraints in the prior distributions. For example, if some covariates have a tendency towards collinearity (\eg, if they are strongly correlated, hence providing redundant information---a common instance in landslide susceptibility and hazard studies), then the prior structure can reduce numerical instabilities and keep the model and its parameters well identifiable from the data. Therefore, the specification of priors is crucial in models where the number of unknown parameters and/or latent random variables to infer is large compared to the observed sample size. 

Here, we will specify different types of prior distributions for distinct model components (\ie, fixed effects and random effects). In particular, we assume that the continuous covariates have been standardised to have mean $0$ and variance $1$, such that the importance of estimated coefficients can be interpreted and compared more easily, and we expect estimated coefficients to be significant if they are moderately large in absolute value. In our models, we then specify a moderately informative normal distribution with mean $0$ and variance $1$ independently for the global intercept and all covariate coefficients. This implies that, \emph{a priori}, the probability of having an absolute covariate coefficient value $\beta$ larger than $2$ is less than $5\%$. However, if the data provide clear evidence that the true coefficient is higher, the final \emph{posterior estimate} of the coefficient can still be much larger without any difficulty. 

As for latent random effects composed of numerous random variables, a general principle is to avoid overly complex models by constructing prior distributions that penalise the model complexity. If the stochastic behavior of such complex model components is not properly controlled by suitably chosen prior distributions, this can lead to overfitting and poor estimation and prediction performances. To penalise model complexity, a strategy is to use informative priors, which shrink complex model components towards simpler reference models, making sure that they can be reliably estimated. This general principle has been formalised recently and leads to an approach based on so-called Penalised Complexity (PC) priors \citep{Simpson.al.2016}. We will make systematic use of PC priors in our implementation. For instance, the reference model for a latent spatial random effect (\ie, the LSE) is simply taken to be its absence, such that the prior distribution of the precision parameter of this effect is designed to avoid overly large spatial variability. In our baseline models, the random effect components are replaced by their reference distribution, \ie, they are absent.

\subsection{Latent Gaussian modelling approach, and the \texttt{R-INLA} library}
\label{sec:R-INLA}

Within the last decade since its publication in 2009, the Integrated Nested Laplace Approximation (INLA) method \citep{Rue2009} has become the most popular tool for Bayesian spatial modelling thanks to its implementation in the \texttt{R-INLA} library \citep[][\href{http://www.r-inla.org/}{http://www.r-inla.org/}]{Lindgren.Rue.2015}  of the statistical software  \texttt{R}  \citep{R}. The general framework of INLA is that of Bayesian latent Gaussian models, which has found widespread interest in a diverse range of applications \citep{Lombardo.etal:2018,Opitz.etal:2018,Krainski.etal:2018,Moraga:2019}. Essentially, the data are assumed to follow a ``well-behaved'' distribution function, and to be conditionally independent given some latent (multivariate) Gaussian random effects. In our context, landslide counts are modelled using the Poisson distribution, whose mean is expressed on the log scale in terms of various fixed effects and latent random effects that are correlated over space and/or time. Each of the covariate coefficients simply has a normal distribution prior, independent from the others. As for hyperparameters, the prior distributions are chosen more specifically depending on the role of the parameter. More details on each of our models are given in the following sections.

The success of INLA relies on the systematic use of random effects with sparse precision (\ie, inverse covariance) matrices within this latent Gaussian modelling framework, coupled with astute analytical and numerical approximation schemes \citep{Illian.al.2012,Rue2009,rue2017bayesian}, which provide exceptional speeds-up for fitting large and complex models compared to more traditional Markov Chain Monte Carlo (MCMC) methods. For details on the theory and practice of \texttt{R-INLA}, we refer the interested reader to the landslide tutorial paper by \citet{lombardo2019numerical} and to \citet{Bakka2018}.
 

\subsection{Baseline LGCP models with fixed effects only}
\label{sec:Baseline}

First, we consider two ``baseline" models, which we call Mod1 and Mod2, where the first one is purely spatial, in the sense that it does not include any information about the structure of time intervals, whereas the second allows for time-interval-specific regression constants. In the model Mod1, we only include the spatially-indexed covariates presented in Section~\ref{sec:covariates} in the log-intensity:
\begin{equation}\label{eq:mod1}
\log\left(\Lambda^{\text{Mod1}}_j(s_i)\right)=\log(|A_i|)+\beta_0+\sum_{k=1}^8 \beta_{k;1}^{\text{morph}} z^{\text{mean}}_k(s_i)+\sum_{k=1}^{8} \beta_{k;2}^{\text{morph}} z^{\text{sd}}_{k}(s_i)+\sum_{k=1}^{13} \beta_k^{\text{them}} z^{\text{prop}}_k(s_i),
\end{equation}
where $j=1,\ldots,6$ indexes the time intervals and each $s_i$, $i=1,\ldots,889$, corresponds to a different SU.
This model comprises $29$ covariate coefficients $\beta$ to be estimated, here separated according to the SU-wise means and standard deviations of morphometric variables obtained from the DEM, with superscript ``$\text{morph}$" in equation~\eqref{eq:mod1}, and the $13$ thematic properties, with superscript ``$\text{them}$" in equation~\eqref{eq:mod1}, expressed through SU-wise proportions (Table~\ref{tab:covariates}). This is a purely spatial model, which assumes that the spatial intensity is the same for each of the 6 temporal intervals (here treated as independent replicates). Moreover, this model does not account for any additional spatially-correlated nor temporally-correlated unobserved effects. 

Model Mod2, with intensity $\Lambda^{\text{Mod2}}_j(s)$, has the following structure: 
\begin{equation}\label{eq:mod2}
\log\left(\Lambda^{\text{Mod2}}_j(s_i)\right)=\log\left(\Lambda^{\text{Mod1}}_j(s_i)\right)+\beta_j
\end{equation}
where $\beta_j$, $j=1,\ldots,6$ are additional time-interval-specific intercepts. We resolve the non-identifiability of $\beta_0$ in \eqref{eq:mod1} and $\beta_j$ in \eqref{eq:mod2} by imposing the sum-to-zero constraint $\sum_{j=1}^6 \beta_j=0$,  such that estimated $\beta_j$-coefficients ($j=1,\ldots,6$) indicate how strongly the overall number of landslides in a time interval deviates from the global average measured through $\beta_0$. As indicated in \S\ref{sec:BayesianModelling}, we assign independent zero-mean Gaussian priors for each regression coefficient. However, unlike the global intercept $\beta_0$ and the covariate coefficients $\beta_{k;1}^{\text{morph}},\beta_{k;2}^{\text{morph}},\beta_{k}^{\text{them}}$ where the prior variance is set to one, the additional intercepts $\beta_j$ specific to each time interval do not have a fixed prior variance; instead, we specify a PC prior for the variance of $\beta_j$ which corresponds to a $50\%$-probability of being below or above $1$, and we then let data decide how strongly the $\beta_j$ values should be allowed to vary between time intervals.

\subsection{Spatial LGCP with replicated spatially structured random effects}
\label{sec:Space}

To extend the baseline models Mod1 and Mod2, we now add a spatial random effect, with $6$ replicates in time, to explain variations in the landslide intensity that cannot be explained by the observed covariates, and we write $\Lambda_j^{\text{Mod3}}(s)$ for the spatial intensity in the $j$-th temporal interval in model Mod3. Our prior assumption is that the spatial random effect should differ between SUs and between different temporal events, but that it tends to be similar for SUs sharing a boundary or being close in space; recall the adjacency graph structure in Figure~\ref{fig:AdjMat}C. 

We first write the general model formula, and we then explain how we encode this prior assumption on spatial dependence for the random effect. The model may be written as 
\begin{equation}\label{eq:mod3}
\log\left(\Lambda^{\text{Mod3}}_j(s_i)\right)=\log\left(\Lambda^{\text{Mod2}}_j(s_i)\right)+W_j^{\text{Mod3}}(s_i), \quad i=1,\ldots,889,\ j=1,\ldots,6,
\end{equation}
where $\Lambda^{\text{Mod2}}_j(s)$ is the baseline intensity of model Mod2 defined in \eqref{eq:mod2}, and where $W_j^{\text{Mod3}}(s)$ is the latent spatial effect (LSE) with 6 replicates, one for each of the temporal intervals.

We use the notation $i_1\sim i_2$ if the SUs $i_1$ and $i_2$ are not the same but share a boundary, and we write $\mathrm{nb}(i)=\{\tilde{i}\mid \tilde{i}\sim i\}$ for the set of all the neighbouring SUs that are adjacent to the $i$-th SU, with $|\mathrm{nb}(i)|$ indicating their number. 
Since the study area is contiguous, we obtain $|\mathrm{nb}(i)| \geq 2$ for all SUs $i=1,\ldots,889$ (Figure~\ref{fig:AdjMat}B). 
The model for $W_j^{\text{Mod3}}(s_i)$ is now specified by the following two conditions:
\begin{enumerate}
    \item The spatial fields $W_{j_1}^{\text{Mod3}}(s)$ and $W_{j_2}^{\text{Mod3}}(s)$ are independent for different times $j_1\not=j_2$.
    \item The value of the spatial random effect $W_j^{\text{Mod3}}(s_i)$ at the $i$-th SU, given all the other values $W_j^{\text{Mod3}}(s_{\tilde{i}})$, $\tilde{i}\neq i$, follows a normal distribution whose mean value corresponds to the mean of the adjacent values, \ie,
    \begin{equation}\label{eq:effspat}
        W_j^{\text{Mod3}}(s_i)\mid\{W_j^{\text{Mod3}}(s_{\tilde{i}});\tilde{i}\neq i\} \sim \mathcal{N}\left(\frac{1}{|\mathrm{nb}(i)|}\sum_{\tilde{i}\sim i}W_j^{\text{Mod3}}(s_{\tilde{i}}),\frac{1}{|\mathrm{nb}(i)|\tau^{\text{Mod3}}}\right),
    \end{equation}
where $\tau^{\text{Mod3}}>0$ is a precision parameter to be estimated, which controls the dependence strength between neighbouring SUs. 
The parameter $\tau^{\text{Mod3}}$ of the conditional spatial distributions in \eqref{eq:effspat} determines the value of the  variance $1/\kappa^{\text{Mod3}}$ of the unconditional effects $W_j^{\text{Mod3}}(s_i)$;  internally, \texttt{R-INLA} implements a parameterisation using the \emph{marginal precision} $\kappa^{\text{Mod3}}$ averaged over all SUs, which may be simpler to interpret in practice. 
\end{enumerate}
The mechanism of this model prescribes that there is less uncertainty about the random effect value in a SU if we know the values in the adjacent SUs. This prior assumption is valid when adjacent slope units have a similar behaviour, \eg, when the sliding surface at depth affects more than one SU, or when the landslide rainfall/seismic trigger has a clear dominating spatial pattern. However, the observed data can also counteract this prior assumption if necessary, \ie, in case two adjacent slope units have strongly different behaviours. This might occur in our study area, \eg, with the common case of bedding planes dipping out of, or nearly parallel to the slope (``cataclinal" slope) in one SU, and dipping into the slope (``anaclinal" slope) in an adjacent SU across a drainage line \citep{marchesini2015assessing,santangelo-MethodAssessment-2015}. The inclusion of the LSE, $W_j^{\text{Mod3}}(s)$, in Mod3 induces spatial dependence within each temporal interval, but keeps the different temporal intervals independent, be they consecutive in time or not.  

\emph{A priori}, we assume that the LSEs have moderate absolute values and are relatively similar between adjacent SUs, such that we construct a prior model whose complexity remains moderate when compared to the baseline Mod2 in \eqref{eq:mod2}. To achieve this, we use a PC prior for the standard deviation parameter $\text{sd}^{\text{Mod3}}=1/\sqrt{\kappa^{\text{Mod3}}}$, which corresponds to an exponential distribution; in mathematical notation, we assume that 
\begin{equation}\label{eq:pcprec}
\text{Pr}(1/\sqrt{\kappa^{\text{Mod3}}}>u)=\exp(-\lambda u),\quad u>0,
\end{equation} 
where $\lambda>0$ is a \emph{penalty rate} to be defined. Here we fix $\lambda$ such that $\text{Pr}(1/\sqrt{\kappa^{\text{Mod3}}}>1)=0.5$, \ie, we give the standard deviation parameter a $50\%$ chance of exceeding $1$ \emph{a priori}. 

\subsection{Temporal LGCP with slope-unit-based temporal random effect}
\label{sec:Time}

Our next model, Mod4, has a similar structure as Mod3 in \eqref{eq:mod3} at first sight, but we now make an assumption about the temporal dependence of SU-based random effects within the same SU, while disregarding any direct spatial relationship between adjacent SUs. Writing $\Lambda_j^{\text{Mod4}}(s)$ for the spatial intensity in the $j$-th temporal interval, the model formula is given as 
\begin{equation}\label{eq:mod4}
\log\left(\Lambda^{\text{Mod4}}_j(s_i)\right)=\log\left(\Lambda^{\text{Mod2}}_j(s_i)\right)+W_{i}^{\text{Mod4}}(t_j), \quad i=1,\ldots,889,\ j=1,\ldots,6,
\end{equation}
where $\Lambda^{\text{Mod2}}_j(s)$ is the baseline intensity of model Mod2 defined in \eqref{eq:mod2}, and where $W_i^{\text{Mod4}}(t)$ is a latent temporal effect (LTE) with $889$ replicates (one for each of the SUs in our study region), defined by the following three conditions: 
\begin{enumerate}
    \item The temporal series $W_{i_1}^{\text{Mod4}}(t_j)$ and $W_{i_1}^{\text{Mod4}}(t_j)$ are independent when considering different SUs $i_1\not=i_2$.
    \item For fixed SU $i$, the value of the temporal random effect $W_i^{\text{Mod4}}(t_j)$ at time $t_j$ with $j>1$, given the value $W_i^{\text{Mod4}}(t_{j-1})$ at the preceding time point $t_{j-1}$, follows a normal distribution with an autoregressive structure, \ie, 
    \begin{equation}\label{eq:temporaldynamics}
        W_i^{\text{Mod4}}(t_j)\mid W_i^{\text{Mod4}}(t_{j-1})\sim \mathcal{N}\left(\beta^{\text{Mod4}} W_i^{\text{Mod4}}(t_{j-1}),\frac{1}{\tau^{\text{Mod4}}}\right),
    \end{equation}
where the temporal autocorrelation parameter $-1<\beta^{\text{Mod4}}<1$ and the precision parameter $\tau^{\text{Mod4}}>0$ have to be estimated. 
\item The value at the first time point $W_i^{\text{Mod4}}(t_1)$ follows a normal distribution with mean $0$ and variance $1/\kappa^{\text{Mod4}}=1/(\tau^{\text{Mod4}}(1-(\beta^{\text{Mod4}})^2))$, \ie,  $W_i^{\text{Mod4}}(t_1)\sim\mathcal{N}(0,1/\kappa^{\text{Mod4}})$.
\end{enumerate}
Equivalently to the conditions 2 and 3 above, we can write $W_i^{\text{Mod4}}(t_j)=\beta^{\text{Mod4}}W_i^{\text{Mod4}}(t_{j-1})+\varepsilon_j$, where the ``innovations" $\varepsilon_j\sim\mathcal{N}(0,1/\tau^{\text{Mod4}})$ are mutually independent and independent of $W_i^{\text{Mod4}}(t_{j-1})$. Under these conditions, the variables $W_i^{\text{Mod4}}(t_j)$ are \emph{a priori} stationary in time with normal distribution $W_i^{\text{Mod4}}(t_j)\sim \mathcal{N}(0,1/\kappa^{\text{Mod4}})$. From an interpretation perspective, the most interesting aspect is to have direct control over the standard deviation $\text{sd}^{\text{Mod4}}=\sqrt{1/\kappa^{\text{Mod4}}}$ and the autoregression parameter $\beta^{\text{Mod4}}$. Here, we proceed as with the spatial model, and we therefore use the same PC prior as in \eqref{eq:pcprec}, setting $\lambda$ such that $\text{sd}^{\text{Mod4}}$ has  $50\%$ chance to exceed the value $1$. When specifying a prior model for $\beta^{\text{Mod4}}$, we assume that consecutive events are only weakly linked to each other, and we implement a PC prior penalizing absolute values of $\beta^{\text{Mod4}}$ close to $1$. Our choice of prior is such that there is a $50\%$ chance for $\beta^{\text{Mod4}}$ to exceed $0.5$ in absolute value. 

\subsection{Space-time LGCP with combined spatial and temporal structures}
\label{sec:space-time}

Finally, considering both the spatial and temporal structures in the data, we construct our most complex model, Mod5, that combines explicit assumptions for the temporal dependence of random effects between consecutive inventories in time, and for spatial dependence based on spatial adjacency relations between SUs (Figure~\ref{fig:AdjMat}C) for contemporaneous landslides, \ie, for landslides in the same time period (Figure~\ref{fig:AggrByTime}).      
Similar to the spatial model Mod3 in \eqref{eq:mod3}, we have a single parameter $\kappa^{\text{Mod5}}>0$ that simultaneously governs the spatial variability and dependence strength of the random effects. Additionally, the parameter $\beta^{\text{Mod5}}$ controls the strength of association between consecutive time intervals, in analogy to the preceding temporal model Mod4 in \eqref{eq:mod4}. Therefore, this space-time model keeps a parsimonious parameterisation with only two hyperparameters $\beta^{\text{Mod5}}$ and $\kappa^{\text{Mod5}}$ to be estimated. Writing now $\Lambda_j^{\text{Mod5}}(s)$ for the spatial intensity in the $j$-th temporal interval in model Mod5, we define
\begin{equation}\label{eq:mod5}
\log\left(\Lambda_j^{\text{Mod5}}(s_i)\right)=\log\left(\Lambda^{\text{Mod2}}_j(s_i)\right)+W^{\text{Mod5}}(s_i,t_j), \quad i=1,\ldots,889,\ j=1,\ldots,6,
\end{equation}
where the latent space-time effect (LSTE), $W^{\text{Mod5}}(s,t)$, now combines the dependence relationships of the purely spatial model Mod3 in \eqref{eq:mod3} and of the purely temporal model Mod4 in \eqref{eq:mod4}. Specifically, we assume the following structure:
\begin{enumerate}
\item The LSTE obeys the following temporal dynamics: 
\begin{equation}
W^{\text{Mod5}}(s_i,t_j)=\beta^{\text{Mod5}}W^{\text{Mod5}}(s_i,t_{j-1})+\varepsilon_j(s_i), \quad -1<\beta^{\text{Mod5}}<1, \quad j>1, 
\end{equation} 
where  the spatial ``innovation fields" $\varepsilon_j(s)$ are mutually independent in time and have the same distribution as the LSE $W^{\text{Mod3}}_1$ in \eqref{eq:effspat}, with $\tau^{\text{Mod3}}$ replaced by $\tau_\varepsilon^{\text{Mod5}}$. We write $\kappa_\varepsilon^{\text{Mod5}}>0$ for the corresponding unconditional precision parameter used internally by \texttt{R-INLA}, in analogy with Model Mod3 in \eqref{eq:mod3}. 
\item The field at the first time point $W^{\text{Mod5}}(s,t_1)$ has the same probability distribution as the LSE $W^{\text{Mod3}}_1$ in \eqref{eq:effspat}, but now we denote the unconditional precision parameter by 
 $1/\kappa^{\text{Mod5}}=1/(\kappa_\varepsilon^{\text{Mod5}}(1-(\beta^{\text{Mod5}})^2))$. This assures that the model is stationary in time for each SU with unconditional precision parameter $\kappa^{\text{Mod5}}$ for all $6$ fields $W^{\text{Mod5}}(s,t_j)$, $j=1,\ldots,6$. 
\end{enumerate}
The prior distribution of the precision parameter 
$\kappa^{\text{Mod5}}$ is fixed as in the spatial model Mod3 in \eqref{eq:mod3}, and the prior of the temporal autocorrelation parameter $\beta^{\text{Mod5}}$ is fixed similarly to the temporal model Mod4 in \eqref{eq:mod4}. 

In the spatio-temporal model Mod5, the data can determine if landslide counts, not well explained by the observed covariates, tend to be similar in space between nearby SUs: \i in case of small mass movements, which separately affect different SUs; or \ii because of single landslide bodies, whose areal extents affect more than one SU. Similarly, the temporal component of the model captures the effect, not explained by the observed covariates, that lead to landslide counts being similar through time between consecutive events for the same SU. In particular, if estimated spatial and temporal dependencies are non-negligible, then the average number of landslides in a SU for a given temporal event is often related to the average number for SUs that are located ``close" in space, and for all the events ``close" in time. Therefore, this model can learn about clustering in space and persistence in time in the structure of the landslide-triggering mechanism. A general understanding of this concept could be practically translated in study areas where, due to orographic conditions, the occurrence of critical precipitation amounts may always tend to arise in the same, relatively confined area in space. The model Mod5 in \eqref{eq:mod5} could capture this trigger structure through spatial dependence (confined area) and temporal dependence (same area for different events), even if no observed data are available for the relevant precipitation events. The latter can be a single trigger in case of event-based inventories or the cumulative effect of several triggers for inventories associated with a large time span. This assumption can be valid when storms have a clear spatial pattern characterised by a transition in precipitation regimes from the ``epicentre" to the peripheries of the cloudburst \citep{Lombardo.etal:2018}. However, the same assumption may not hold in our study area because of its size (less than $10$ $\times$ $10 km^2$) and the absence of significant orographic gradients. An alternative explanation that may relate more closely to our study case could be that the spatial dependence captures the unknown effect of land use or land cover, driven, \eg, by changing economy and agricultural practices, or the dependence induced by large landslides destabilizing more than one SU.

\subsection{Implementation and model validation}
\label{sec:Implementation}

To assess the models' performance and their interpretability from an explanatory perspective, we chose to implement an initial modelling stage where we fitted the baseline LGCPs (Mod1 and Mod2) and the three extensions featuring the latent fields (Mod3, Mod4, and Mod5) using $100\%$ of the dataset. Subsequently, we assessed the predictive performance of each model. To do this, we implemented two separate cross-validation (CV) schemes. The first approach is a spatial 10-fold CV, whereby we extract 10\% of the dataset for testing and leave out the complementary 90\% for fitting. The proportion of held-out data may seem relatively small in comparison to commonly used values in the literature (\eg, $30\%$ to $50\%$, \citet{reichenbach2018}), but we consider it as a sensible value to allow the model to learn about spatial structure in the training data. We constrained the random extractions to avoid any SU to be sampled more than once, and we used the same SUs across time intervals. In other words, the combination of the ten CV complementary subsets reproduces the original dataset. The second CV approach is a temporal  ``leave-one-out" procedure whereby six models are fitted, each one calibrated on five time intervals and tested on the remaining one, regardless of the temporal position of the time periods used, \ie, the fitted model does not account for possible landslide heritage effects \citep{samia-CharacterizationQuantification-2017,samia-LandslidesFollow-2017}. 

We assessed the accuracy of the estimated landslide intensities both for the fit and cross-validation procedures by comparing observed and predicted landslide counts for each model, and for each temporal interval. Here, predicted counts for SU $i$ and time interval $j$ are defined as the posterior mean $\hat{\Lambda}_{j}(s_i)$ of the corresponding intensity model $\Lambda_j(s_i)$. We also maintained a link to the ``traditional'' landslide susceptibility, \ie, the probability of observing one or more landslides in a given SU at a given time \citep{chung-ProbabilisticPrediction-1999,guzzetti1999landslidehazard,van1999}. Thanks to the Poisson distribution of landslide counts, the fitted intensity $\hat{\Lambda}_j(s_i)$ for the $i$-th SU and $j$-th temporal interval can be converted into the landslide susceptibility $S_j(s_i)$ via the following equation:
\begin{align}
S_j(s_i) = 1-e^{-\hat{\Lambda}_j(s_i)}.
\label{eq:conversion}
\end{align}


Hence, we computed performance metrics for models Mod1 to Mod5, both in terms of landslide intensity and susceptibility, for each of the two CV schemes. 

\subsection{Classification of intensity and susceptibility estimates}
\label{sec:Classification}

In the statistically-based landslide susceptibility literature, there is no agreement on how to reclassify and show in map form the continuous spectrum of probability values that result from a classification model into few meaningful classes \citep{reichenbach2018}; a fundamental step for a susceptibility zonation to be used in practical applications \citep{guzzetti-ComparingLandslide-2000,galli-ComparingLandslide-2008}. The simplest option is to divide the entire probability range $[0,1]$ into two classes of predicted ``stable" and ``unstable" conditions; with the stable conditions predicted not to have landslides, and the unstable conditions predicted to have landslides. Even for this simple twofold division, several approaches are found in the literature with authors arbitrarily setting the probability cutoff. For presence-absence balanced datasets, examples exist where the cutoff is set to $0.5$ \citep{dai2002} without providing an explanation \citep[\eg,][]{suzen2012}, or because it corresponds to the mean value between the two extremes of the theoretical probability range \citep[\eg,][]{lombardo2016b}. The approach is problematic, because it sets the cutoff where the model is most uncertain \citep{rossi2010optimal,reichenbach2018}. \citet{Frattini2010} pointed out that the choice of a cutoff value depends on the proportion of the presence-absence cases, leading to (much) smaller probability cutoff in datasets with a larger prevalence. For instance, \citet{vanDH2006} reported an upper cutoff of 0.0012 for the high susceptibility class in a study case in the Flemish Ardennes where landslides were rare. In an attempt to address the issue, \citet{Camilo2017} proposed to select a cutoff value that maximises the model accuracy, obtained testing thousands of probability values from the estimated susceptibility.

The problem becomes even more complex when the classification scheme involves more than two classes, typically three to seven \citep{reichenbach2018}. A number of authors have used quantiles to segment the continuous probability estimates into discrete susceptibility classes, \eg, from ``low" to ``high" susceptibility. As an example, \citet{lombardo2018presenting} used $2.5$, $25$, $50$, $75$, and $97.5$ percentiles (\ie, where the three central values are motivated from the classical ``boxplot''). Other authors segmented their probability estimates by counting the number of observed landslides in each probability class. As an example, \citet{petschko.etal:2014} counted the number of landslides in each susceptibility ``bin", and set probability thresholds corresponding to 5\%, 25\%, and 70\% of the total observed landslides. An alternative approach was proposed by \citet{chung2003}, who ranked their probability estimates based on an ``effectiveness ratio", \ie, the ratio between the proportion of total landslide area, $A_{\text{LT}}$ in each susceptibility class and the proportion of the susceptibility class in the study area. \citet{Guzzetti2006b} adopted this approach to show the result of their susceptibility zonation for the Collazzone study area. The mentioned approaches have their rationale, but examples exist in the literature for which there is no clear justification for the adopted classification system, especially when the aim is to produce a predictive map rather than measuring the model performance. For instance, \citet{ayalew2005application} reported that slicing the probability domain into equally-spaced bins was not optimal in their case, and suggested using bins equal to the standard deviation of the probability. Similarly, \citet{pourghasemi2012} used the natural break tool in ArcGIS\textsuperscript{\textregistered} without providing an explanation for this choice. This is the current, controversial, and unclear state of play in the landslide susceptibility literature. We further note that all these methods fail to recognise that probability values near $0.5$ reveal the inability of the model to determine if a mapping unit (\eg, a SU) is stable or unstable, and do not represent  ``medium" or ``intermediate" susceptibility conditions, or levels \citep{rossi2010optimal,reichenbach2018}.

In this work, we propose an innovative strategy to classify and rank the intensity and the susceptibility estimates, and theirs associated uncertainties, provided by our five models. 
We adopt the following four-classes ranking scheme: 

\begin{itemize}
    \item Clearly stable SUs: $\text{intensity} \leq 0.05$, equivalent to $\text{susceptibility} \leq 0.05$. This class corresponds to an intensity/susceptibility range, where the model predicts a lack of landslide occurrences with high probability. More precisely, for the SUs in this class, the probability of having no landslide is estimated to be more than $95\%$.
    \item Uncertain Type 1: $0.05 < \text{intensity} \leq 1$, equivalent to $0.05 < \text{susceptibility} \leq 0.63$. This class characterises SUs that have a probability of landslide occurrence being greater than $5\%$, while their expected landslide count is less than one (on average).
    \item Uncertain Type 2: $1 < \text{intensity} \leq 3$, equivalent to $0.63 < \text{susceptibility} \leq 0.95$. This class characterises SUs that have a probability of landslide occurrence being less than $95\%$, while their expected landslide count is more than one (on average).
    \item Clearly unstable SUs: $\text{intensity} > 3$, equivalent to $\text{susceptibility} > 0.95$. This class corresponds to an intensity/susceptibility range, where the model predicts landslide occurrences with high probability. More precisely, for the SUs in this class, the probability of having at least one landslide is estimated to be more than $95\%$, and the expected landslide count is more than $3$ per SU.
\end{itemize}


This new classification is applied and discussed below in \S\ref{sec:best-model}.

\section{Results}
\label{sec:results}

In this section, we present the results of our modelling effort. First, in \S\ref{sec:BaselineResults}, we outline the results of our simplest, baseline model Mod1 in \eqref{eq:mod1}, showing the model intensity and susceptibility estimates. 
Model Mod1 is the closest LGCP counterpart to more ``traditional'' susceptibility models, and it represents a good reference against which to compare the more advanced models, which we do in \S\ref{sec:AdvancedResults}. This is followed by the presentation of the models' fitting performance (\S\ref{sec:Fit}), of the latent temporal (\S\ref{sec:TimeEffects}) and linear covariates (\S\ref{sec:CovariateEffects}) effects, of the models prediction skills (\S\ref{sec:CV}), and of what we consider our best intensity--susceptibility model (Mod5) (\S\ref{sec:best-model}). Lastly, in \S\ref{sec:CR}, we provide information on the computational requirements to fit our models.

\subsection{Baseline intensity and susceptibility estimates}
\label{sec:BaselineResults}

\begin{figure}[!htbp]
	\centering
	\includegraphics[width=1\linewidth]{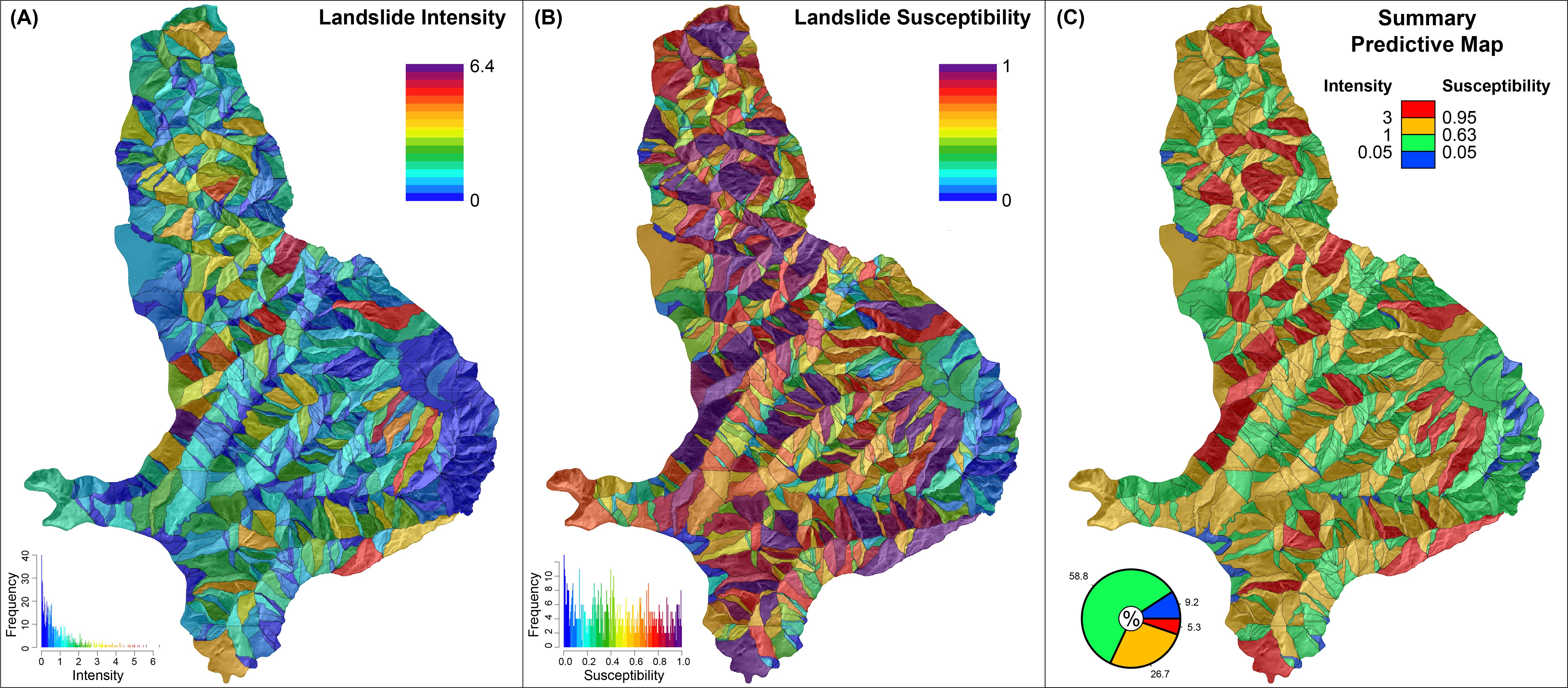}
	\caption {Maps show (A) the fitted baseline landslide intensity, (B) susceptibility, and (C) unified reclassification estimated based on model Mod1 constructed  using the morphometric, geologic, and bedding attitude variables listed in Table~\ref{tab:covariates}. In (A) and (B), histograms show the frequency of the modelled intensity and susceptibility values. In (C), pie-chart shows the percentage of SUs falling into one of the four considered classes.}
	\label{fig:BaselineMap}
\end{figure}

For each of the 889 SUs that partition our study area, Figure~\ref{fig:BaselineMap} shows, in map form, the fitted estimates obtained by model Mod1 exploiting the morphometric and thematic covariates listed in Table~\ref{tab:covariates}, and without considering any spatial or temporal latent dependency in the data.
Figure~\ref{fig:BaselineMap}A shows the spatial distribution of the landslide intensity, \ie, of the number of landslides that, on average, can be expected in each SU, based on the used covariates. More precisely, the map shows the temporal average, from T1 to T6 (see Figure~\ref{fig:AggrByTime}), of the modelled intensities for each time interval, \ie, their posterior means. We note that model Mod1 estimated up to $6.4$ landslides per SU; a Figure~that is significantly lower than the maximum number of landslides (25, see Figure~\ref{fig:AggrByTime}) in a SU in the multi-temporal inventory (Figure~\ref{fig:StudyArea}). This is a limitation of model Mod1. Figure~\ref{fig:BaselineMap}B shows the spatial distribution of landslide susceptibility, \ie, the propensity (proneness) of each SU to generate landslides, based on the considered local terrain conditions (Table~\ref{tab:covariates}). In the map, the susceptibility estimates were obtained from the intensity estimates of Figure~\ref{fig:BaselineMap}A using equation~\eqref{eq:conversion}. In the intensity (Figure~\ref{fig:BaselineMap}A) and the susceptibility (Figure~\ref{fig:BaselineMap}B) maps, the histograms show the frequency distributions of the modelled intensity and susceptibility estimates. Visual comparison of the two histograms reveals that the distributions are significantly different; with the distribution of the intensity estimates positively skewed, and the distribution for the corresponding susceptibility estimates more uniform. This was expected, as the the number of SUs that can generate a large or very large number of landslides is limited in the study area, whereas the number of SUs that can generate landslides, \ie, that are potentially ``susceptible" to landslides, is large and geographically distributed.

Lastly, Figure~\ref{fig:BaselineMap}C portrays a joint landslide intensity--susceptibility map prepared adopting the four-class ranking scheme proposed in \S\ref{sec:Classification}, which gives a combined view of the other two maps. In the combined map, out of the 889 SUs, 82 (9.2\%) are classified as ``clearly stable" (blue), and 47 (5.3\%) as ``clearly unstable" (red). Overall, the terrain estimated to be ``clearly stable" covers 1.9 $km^2$, 2.4\% of the study area, and the terrain estimated to be ``clearly unstable" covers 13.3 $km^2$, 16.9\% of the study area. In the ``clearly stable" SUs, the estimated landslide intensity is expected to be $\leq 0.05$, and the corresponding susceptibility also $\leq 0.05$, \ie, very low. Conversely, in the ``clearly unstable" SUs, the landslide intensity is expected to be $> 3$, \ie, three or more expected landslides, and the susceptibility is $> 0.95$, \ie, very high. Further inspection of Figure~\ref{fig:BaselineMap}C reveals that the majority of the SUs (58.8\%), covering 28.5 $km^2$, 36.1\% of the study area, is considered of  ``Uncertain Type 1", followed by 26.7\% of ``Uncertain Type 2", covering 35.2 $km^2$ or 44.7\% of the Collazzone area. In these areas, on average, less (respectively more) than one landslide is expected in each SU. 

\subsection{Advanced intensity and susceptibility estimates}
\label{sec:AdvancedResults}

\begin{figure}[!h]
	\centering
	\includegraphics[width=\linewidth]{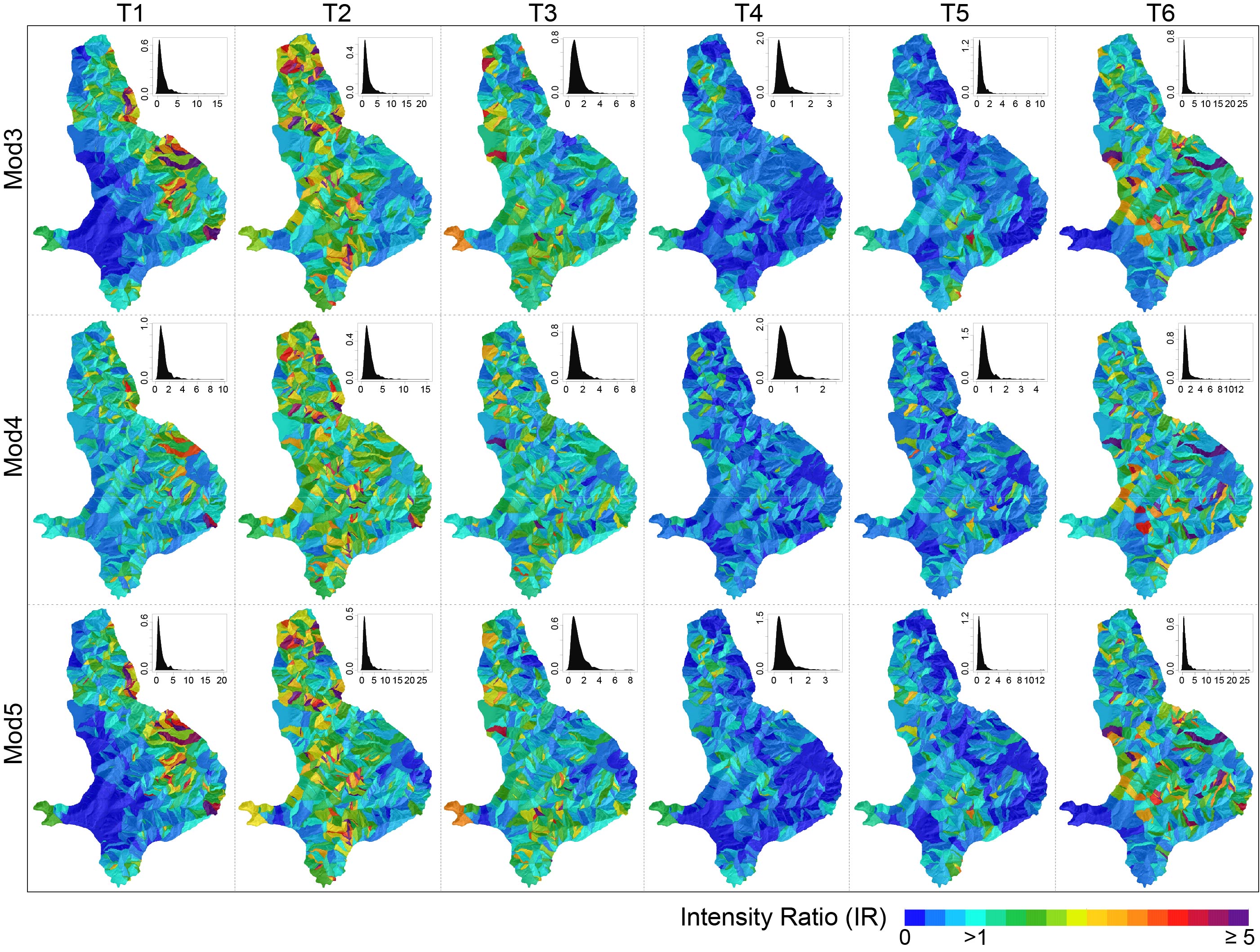}
	\caption{Estimated intensity ratios (IRs) for the three models Mod3, Mod4, and Mod5, compared to the baseline model Mod1, for each temporal interval (T1 to T6) 
	(see Figure~\ref{fig:AggrByTime}). The shown values, $\hat{\Lambda}^{\text{Mod3}}_j(s_i)/\hat{\Lambda}^{\text{Mod1}}_j(s_i)$, $j=1,\ldots,6$, express the factor by which the intensity estimates for model Mod1 (shown in Figure~\ref{fig:BaselineMap}A) have to be multiplied to get the estimated intensity for models Mod3, Mod4, and Mod 5, respectively. Color bar is uniform but limited to values greater than or equal to a factor of five for graphical purposes. Small graphs show density of intensity ratios (IRs) for each model and temporal interval. Note that x-axes and y-axes in the individual graphs cover different ranges.	See text for explanation.} 
	\label{fig:Fig5}
\end{figure}

\begin{figure}[!h]
	\centering
	\includegraphics[width=\linewidth]{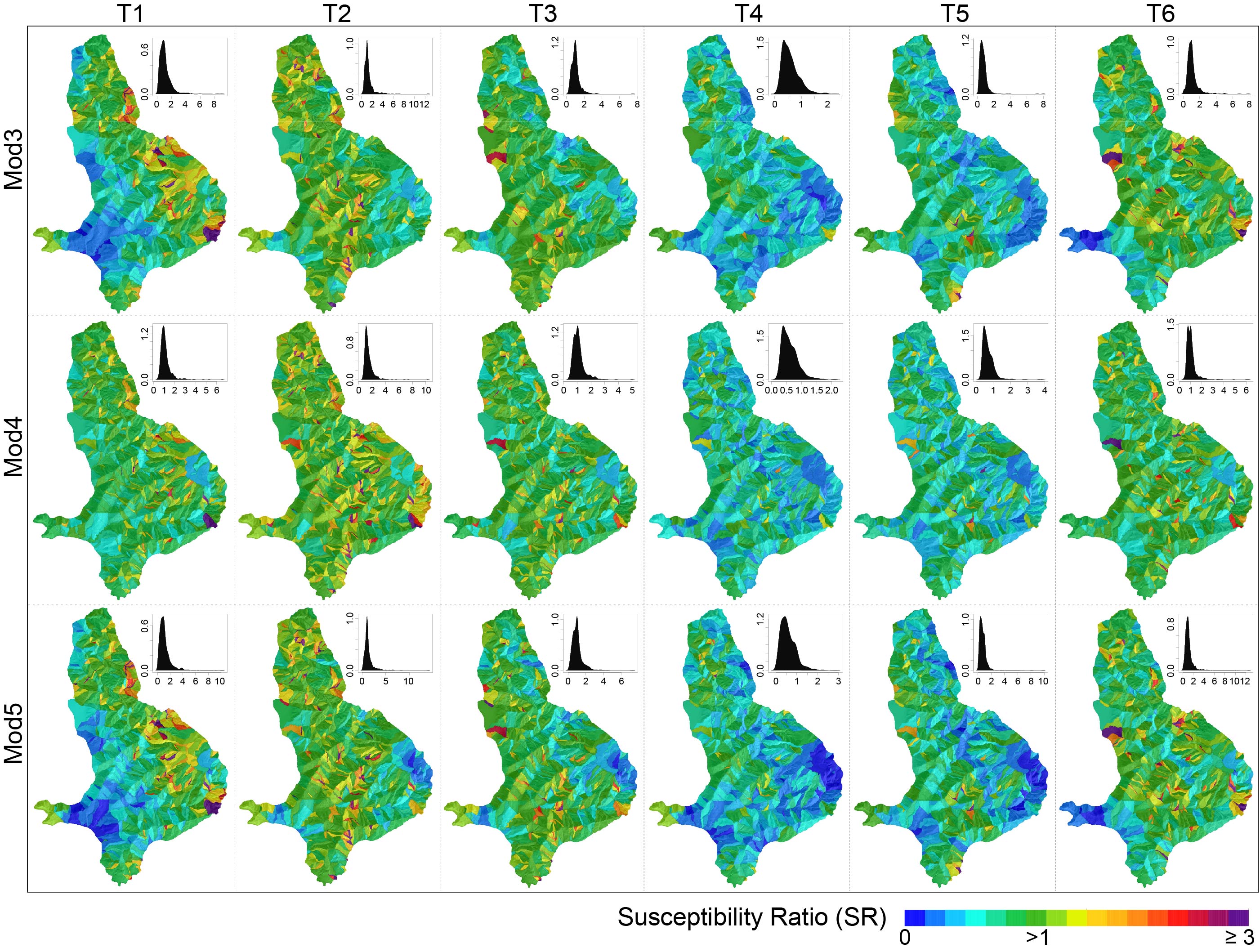}
	\caption{Estimated susceptibility ratios (SRs) for the three advanced models Mod3, Mod4, and Mod5, compared to the baseline model Mod1, computed for each temporal interval (T1 to T6). The reported values, $\hat{\Lambda}^{\text{Mod3}}_j(s_i)/\hat{\Lambda}^{\text{Mod1}}_j(s_i)$, $j=1,\ldots,6$, express the factor by which the susceptibility estimates for Mod1 (shown in Figure~\ref{fig:BaselineMap}B) have to be multiplied to get the estimated intensity for Mod3, Mod4, and Mod 5, respectively. Color bar is uniform but limited to values greater than or equal to a factor of five, for graphical purposes. Small graphs show density of susceptibility ratios (SRs) for each model and temporal interval. Note that x-axes and y-axes in the individual graphs cover different ranges. See text for explanation.}
	\label{fig:Fig6}
\end{figure}

To highlight the higher flexibility and the better performance of the more advanced models Mod3, Mod4, and Mod5, featuring spatial, temporal, and spatio-temporal latent effects, respectively, we show in Figure~\ref{fig:Fig5} the relative intensity maps, \ie, the ratio of the intensity estimates obtained for each time interval (T1 to T6) obtained by models Mod3, Mod4, and Mod5, to the baseline model Mod1, and in Figure~\ref{fig:Fig6} the corresponding relative susceptibility maps, showing the ratio of the susceptibility estimates obtained by the three advanced models to the baseline model Mod1.

Interestingly, the values portrayed in the maps shown in the two Figures~\ref{fig:Fig5} and~\ref{fig:Fig6} represent the factors  by which the intensity and susceptibility baseline estimates should be multiplied to account for the  space, time, and space-time dependencies. More precisely, the maps in the 18 panels of Figure~\ref{fig:Fig5} show $\hat{\Lambda}^{\text{Mod3}}_j(s_i)/\hat{\Lambda}^{\text{Mod1}}_j(s_i)$, $\hat{\Lambda}^{\text{Mod4}}_j(s_i)/\hat{\Lambda}^{\text{Mod1}}_j(s_i)$ and $\hat{\Lambda}^{\text{Mod5}}_j(s_i)/\hat{\Lambda}^{\text{Mod1}}_j(s_i)$, estimated from each respective model for the different time intervals $j=1,\ldots,6$. Similarly, using the intensity--susceptibility conversion equation \eqref{eq:conversion}, Figure~\ref{fig:Fig6} portrays $[1-\exp\{-\hat{\Lambda}^{\text{Mod3}}_j(s_i)\}]/[1-\exp\{-\hat{\Lambda}^{\text{Mod1}}_j(s_i)\}]$, $[1-\exp\{-\hat{\Lambda}^{\text{Mod4}}_j(s_i)\}]/[1-\exp\{-\hat{\Lambda}^{\text{Mod1}}_j(s_i)\}]$, and $[1-\exp\{-\hat{\Lambda}^{\text{Mod5}}_j(s_i)\}]/[1-\exp\{-\hat{\Lambda}^{\text{Mod1}}_j(s_i)\}]$, $j=1,\ldots,6$, respectively. In this way, the different maps highlight the similarities and the differences in intensity and susceptibility with respect to model Mod1, facilitating the interpretation of the performance of the advanced models Mod3, Mod4, and Mod5. To facilitate the visual comparison of the patterns of the estimated latent effects, when preparing the maps we choose different colour bars valid for each map in the two Figures \ref{fig:Fig5} and \ref{fig:Fig6}. This was obtained saturating the colour scales at a factor of 5 for the Intensity ratios (IRs), and at a factor of 3 for the Susceptibility ratios (SRs). To further help the comparison, in each map we show the probability density distributions of the estimated values.

For each of the advanced models Mod3, Mod4, and Mod5, the estimated intensity and susceptibility ratios strongly differ from one, in several areas and time intervals, suggesting that the inclusion of latent random effects in these complex models is necessary to capture the large variations in intensity and susceptibility across space and time. In other words, these variations cannot be explained solely by the available covariates included in model Mod1. The higher flexibility of the random effect models may thus improve the goodness-of-fit and prediction skills. This will be investigated in more detail in \S\ref{sec:Fit} and \S\ref{sec:CV}. We further note a clear resemblance between the intensity and susceptibility ratios of the spatial and spatio-temporal models, namely Mod3 and Mod5. This hints at a greater effect of the spatial dimension with respect to the temporal dimension in explaining the known distribution of landslides (Figure~\ref{fig:StudyArea}).

\subsection{Fitting models performance}
\label{sec:Fit}

\begin{figure}[!h]
	\centering
	\includegraphics[width=0.7\textwidth]{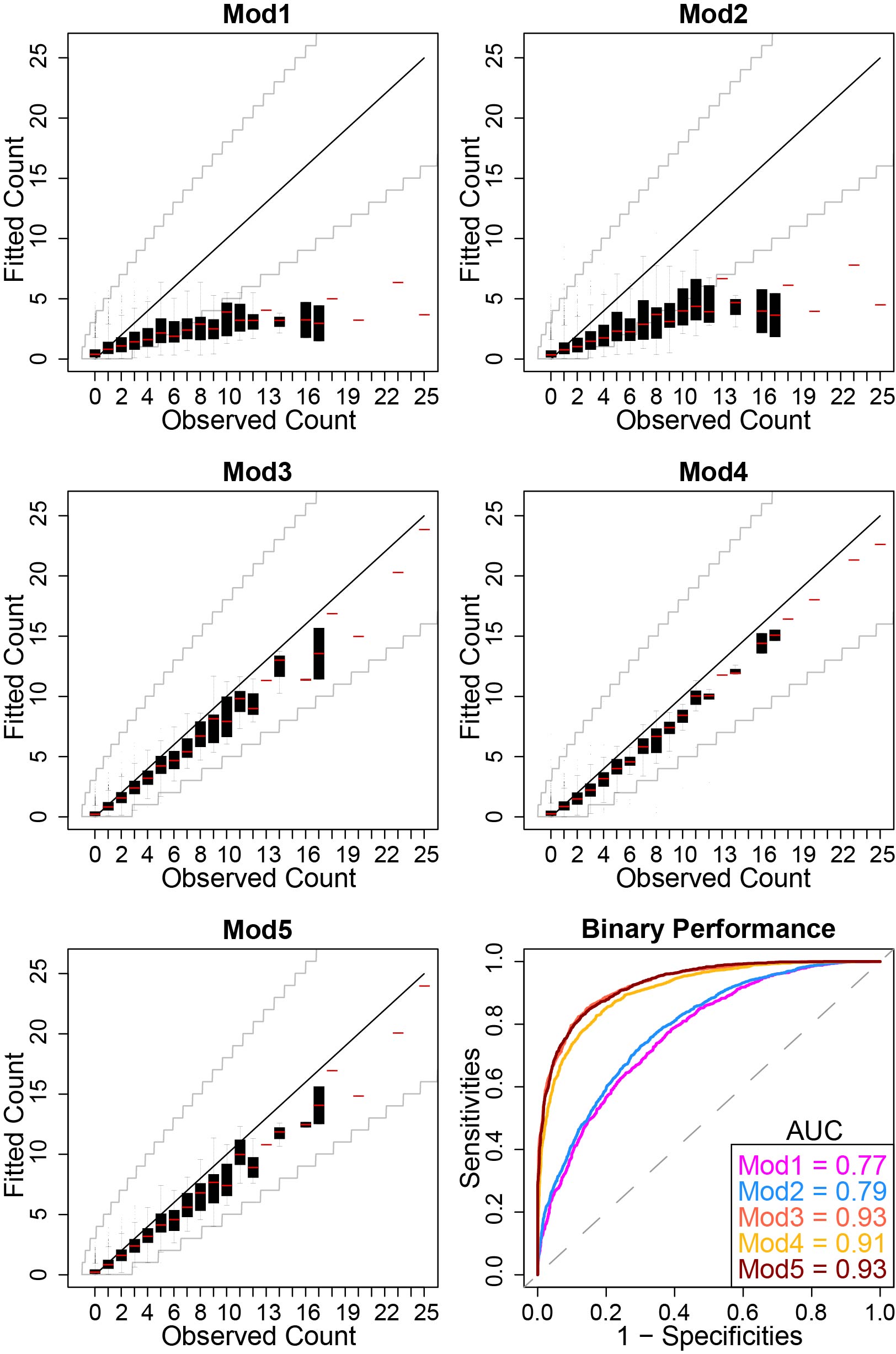}
	\caption{Within-sample performance summary of each model. Fitted landslide counts plotted against observed counts for the five models. For each observed count, we display boxplots of fitted counts to summarise their distribution across all SUs. The grey lines correspond to a 95\% interval for an exact Poisson distribution with mean between 0 and 25. Bottom right panel shows the results in a susceptibility framework, plotting the ROC curve for each model and summarising the within-sample goodness-of-fit performance with the corresponding AUC values. The higher the AUC value, the better the model fit.}
	\label{fig:CountFit}
\end{figure}

For each SU in the study area, Figure~\ref{fig:CountFit} compares the observed to the estimated landslide counts, for the five models (Mod1 to Mod5). It also shows the susceptibility estimates calculated using equation \eqref{eq:conversion}, to facilitate comparison with traditional landslide predictive studies. Visual inspection of Figure~\ref{fig:CountFit} reveals a clear pattern where the baseline models, Mod1 and Mod2, strongly underestimate the observed counts larger than one, whereas they often tend to largely overestimate the zeros (\ie, no landslides). By contrast, the advanced models, Mod3, Mod4 and Mod5 that account for space, time, and space-time dependencies, respectively, significantly improve the goodness-of-fit, with points aligned closer to the diagonal, the latter corresponding to a perfect fit. Similarly, the Receiver Operating Characteristic (ROC) curve and the Area Under the Curve (AUC) computed for the five models show a similar situation where Mod1 is the weakest, followed by Mod2, whose performance is slightly better due to the contribution of the multiple temporal intercept. Overall, Mod3, Mod4, and Mod5 perform much better, improving the two baseline results from acceptable to outstanding, according to \citet{hosmer2000}. 

\subsection{Temporal effects}
\label{sec:TimeEffects}

To investigate the temporal dynamics driving landslide occurrence in our study area, we now focus on model Mod4 in \eqref{eq:temporaldynamics}, in which we decompose the temporal effect into global multiple intercepts (with one coefficient for each time interval), assumed to be \emph{a priori} independent across time, and latent temporal effects (LTEs) for each SU, assumed to be driven by an autoregressive temporal dependence structure \citep{Blangiardo.Cameletti.2015,Opitz.2017}. While the multiple intercepts are constant in space and capture abrupt changes in the overall landslide intensity over time (\eg, due to triggers of different magnitudes), the LTE is designed to capture local, SU-specific changes that are smoother in time, and we thus make the assumption that the LTE carries information about ``clustering" and ``repellency" effects in each SU. 

\begin{figure}[!t]
	\centering
	\includegraphics[width=0.5\textwidth]{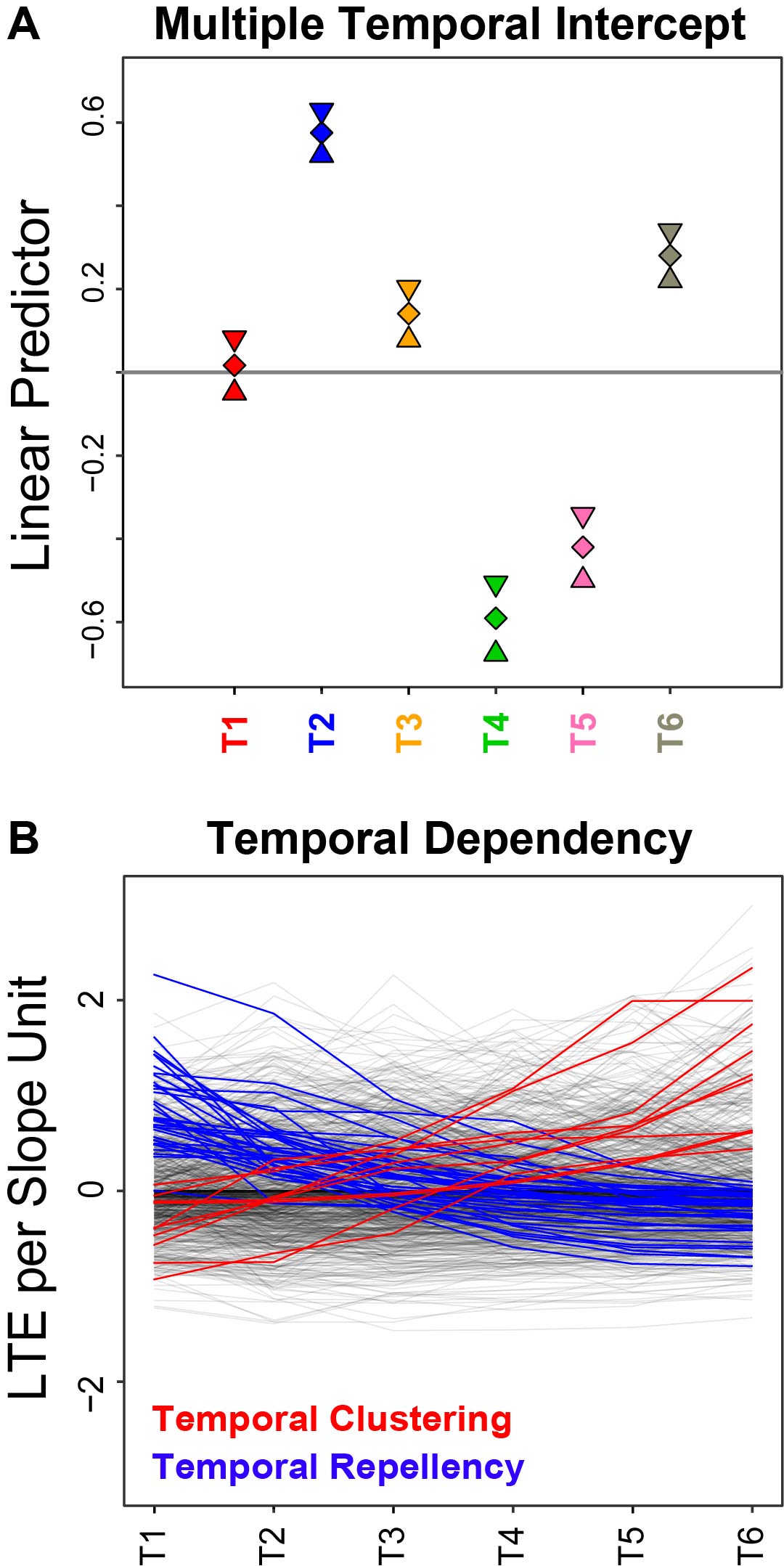}
	\caption{
	Panel A shows  multiple intercepts for each time interval, from T1 to T6, for  model Mod4. 
	Diamonds show posterior intercept means; 
	triangles show pointwise 97.5 and 2.5 posterior percentiles bracketing 95\% pointwise credible intervals. 
	Panel B shows posterior mean of the latent temporal effects (LTE) estimated for all $889$ slope units in the study area. Strict temporal clustering effects (with the LTE increasing monotonically) are shown in red (10 SUs), and strict temporal repellency effects (with the LTE decreasing monotonically) are shown in blue (34 SUs). Grey lines are the remaining SUs for which a strict classification cannot be made (845 SUs). 
	}
\label{fig:TimeDependence}
\end{figure}

The plot in Figure~\ref{fig:TimeDependence}A shows the posterior distribution of the multiple intercepts. 
Inspection of the plot reveals a sudden increase of the multiple intercept during T2 (1941-1954). This is the result of a severe regional rainfall event that hit Central Italy in December 1937 \citep{reichenbach1998}, resulting in numerous landslides in the Collazzone area, which was captured in the multi-temporal inventory interpreting aerial photographs taken in 1941. In this sense, the multiple intercepts carry the strength of the overall effect of the triggers in the six time periods. Conversely, the LTE captures more localised effects in each SU, estimating the relation between landslide counts in a given time period and the number of landslides in the following time period. 

In Figure~\ref{fig:TimeDependence}B, we show the temporal evolution of the posterior means of the LTE for the $889$ SUs in the study area, in a single plot. Inspection of the plot reveals that most of the SUs (845, \ie, 95.0\%) exhibit erratic temporal trends, with LTE increasing or decreasing ``randomly" in time (grey lines in Figure~\ref{fig:TimeDependence}B). Closer inspection of the plot reveals that \i a small number of SUs (34, \ie, 3.8\%) exhibits a monotonically decreasing trend of the LTE (blue lines in Figure~\ref{fig:TimeDependence}B), and \ii an even smaller number of SUs (10, \ie, 1.2\%) exhibit a monotonically increasing trend of the LTE (red lines in Figure~\ref{fig:TimeDependence}B). From a geomorphological perspective, the first group encompasses SUs characterised by landslide temporal ``repellency", where the presence of a landslide in a time period has hampered the occurrence of new landslides in the future periods in the same SU, whereas  the second group encompasses SUs characterised by temporal ``clustering", where new landslides have continually followed previous landslides in the same SU, for the entire considered period (T1--T6). The later result agrees with the findings of \citet{Samia2018} who have identified a ``temporal path dependency" of new landslides on pre-existing landslides in the Collazzone area. Interestingly, the temporal response of landslide path dependency identified by \citet{Samia2018} disappears after about 10--15 year in the study area, \ie, within most of the time periods considered in this study. This explains the reduced number of SUs characterised by distinct temporal ``clustering" found in this study.

\subsection{Covariates effects}
\label{sec:CovariateEffects}

\begin{figure}[!htbp]
	\centering
	\includegraphics[width=0.792\linewidth]{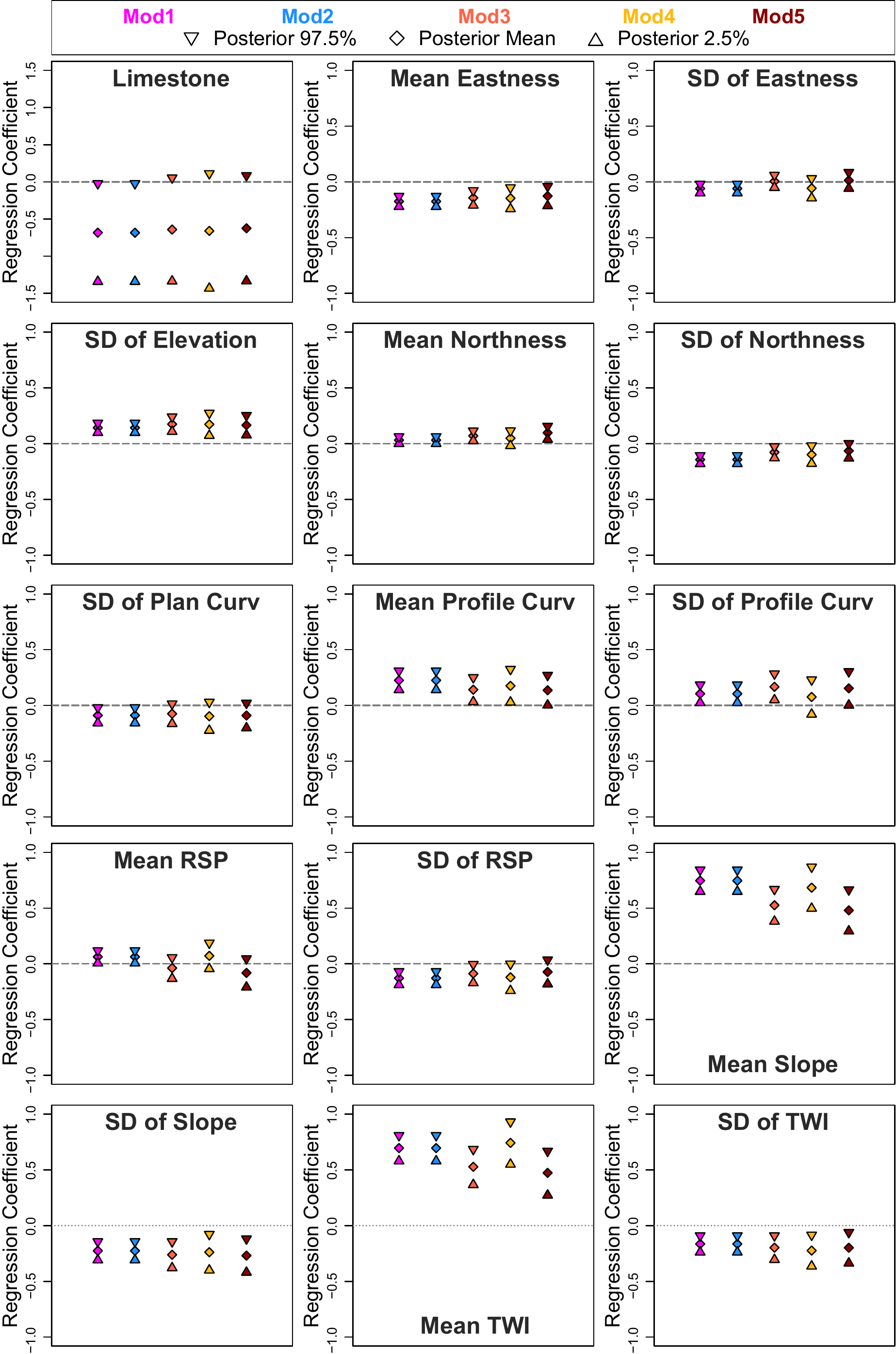}
	\caption{Regression coefficients of all covariates, whose coefficients were significant in at least one of the fitted models, Mod1 (pink), Mod2 (blue), Mod3 (orange), Mod4 (yellow), and Mod5 (brown). Diamonds show posterior means; triangles show pointwise 97.5 and 2.5 posterior percentiles 
	bracketing 95\% pointwise credible intervals.}
	\label{fig:Coefficients}
\end{figure}

Figure~\ref{fig:Coefficients} shows a summary of the posterior distribution of all the estimated regression coefficients that appeared to be significant in at least one of the five models. In the plots, we also show the estimated coefficients for Mod1 and Mod2 to highlight an issue common to all regression models in which residual dependence is not accounted for appropriately. In fact, the 95\% credible interval of the regression coefficients estimated for Mod1 and Mod2 (with no latent effects included) is narrower than the credible intervals for the models with the LSE (orange), LTE (yellow) and LSTE (brown). This is a result of the model structure, where the simpler models (Mod1 and Mod2) are overconfident of the information carried by the observations, whereas the models that incorporate spatial (Mod3), temporal (Mod4) and spatio-temporal (Mod5) dependencies estimate more realistic credible intervals. In our case, the differences are small, and the pattern shows that the five models assign analogous posterior mean values to each covariate, both in amplitude and sign. We consider this an evidence of the overall goodness-of-fit of the different models.

Determining whether a predictive model is useful in practice depends, among other factors, on the interpretability of the estimated covariates' effects. Out of all the 29 covariates used to construct the models (see Table~\ref{tab:covariates}), 15 were significant in at least one model (see Figure~\ref{fig:Coefficients}). As the estimated regression coefficients were similar across the five models, we now provide a unique interpretation. The \emph{Mean Slope} and \emph{Mean TWI} variables gave the strongest contribution to the models, with coefficients much larger (in absolute value) than the coefficients of all other covariates. Both variables contributed to increase the landslide intensity (hence the susceptibility) in all the SUs. From a geomorphological perspective, terrain slope controls the balance of the retaining and the destabilising forces acting on a slope \citep{taylor1948,wu-DistributedSlope-1995,Donnarumma2013}, and in many areas and for landslides of the slide type \citep{Hungr2014} like the one prevalent in the Collazzone study area, terrain slope and its derivatives (\eg, mean slope, slope range, standard deviation of slope) are known to be positively (albeit not necessarily linearly) correlated to the presence and abundance of landslides, and hence to landslide susceptibility \citep{Carrara1991,carrara-GISTechnology-1995,fabbri-PredictionFuture-2003,budimir-SystematicReview-2015,lombardo2018presenting}.
\emph{TWI} measures the ability of a given area to retain surface water as a function of the terrain gradient and the upslope contributing area, favouring infiltration and the increase of the pore water pressure at depth, and, hence, slope instability \citep[\eg,][]{yilmaz2009landslide,cama2017improving}. 

Curvature primarily controls convergence and divergence of overland flows which is often linked to slope stability \citep[\eg,][]{Ohlmacher2007}. Here, laterally-concave \emph{Planar Curvature} is estimated to contribute to landslide-prone conditions, whereas mean upwardly-concave \emph{Profile Curvature} conditions and their variability within a SU increase the expected number of landslides. This effect is exacerbated by the standard deviation of \emph{Elevation} which is a proxy for terrain roughness. Our five models concurred that an increase in the standard deviation of \emph{Elevation} within a SU contributed to an increase in the estimated number of landslides (\ie, a larger intensity), and hence to a larger susceptibility. The mean \emph{Relative Slope Position (RSP)} was significant only for Mod1 and Mod2, although larger \emph{RSP} values contributed to increasing the estimated intensity, for all five models. The \emph{RSP} is a continuous index which essentially assigns 0 to lowland and flat areas, and up to 1 to mountain tops. Thus, a positive regression coefficient suggests that SUs located mostly in the high portion of the local topography are more prone to landslides than SUs located chiefly in the lower part of the local topography.  

\begin{figure}[!htbp]
	\centering
	\includegraphics[width=0.55\linewidth]{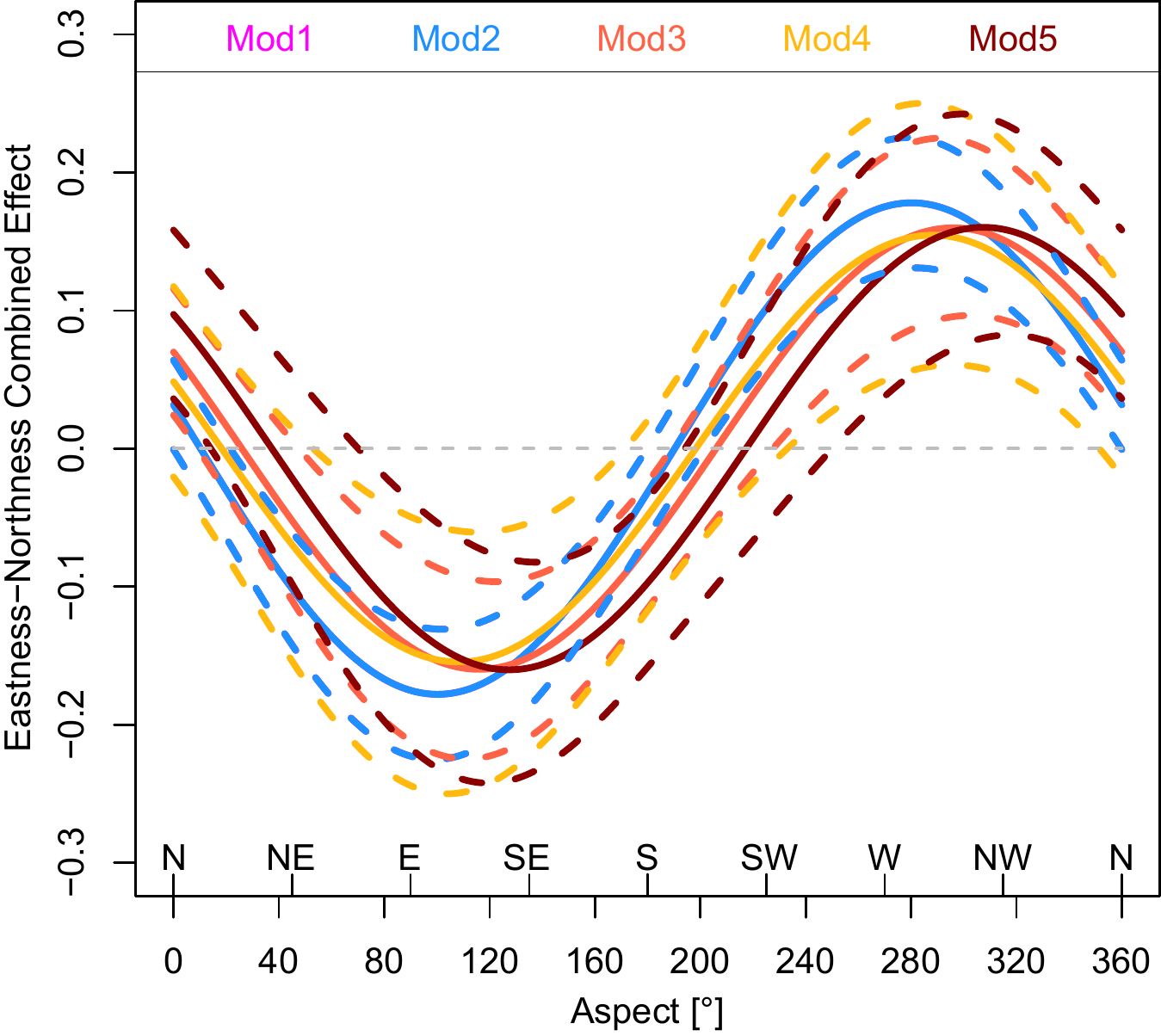}
	\caption{Estimated effect (solid curves) of terrain aspect for models Mod1 (pink), Mod2 (blue), Mod3 (orange), Mod4 (yellow), and Mod5 (brown), obtained by combining the effects of \emph{Eastness} and \emph{Northness}. The plot shows $\beta_{\text{Eastness}}\cos(\theta)+\beta_{\text{Northness}}\sin(\theta)$ as a function of  $\theta\in[0,360]^\circ$, where $\beta_{\text{Eastness}}$ and $\beta_{\text{Northness}}$ here denote (with some abuse of notation) the \emph{Eastness} and \emph{Northness} coefficients, respectively, estimated from each model. Dashed lines show corresponding $95\%$ credible bands.
	Curves for models Mod1 and Mod2 cannot be distinguished.}
	\label{fig:AspectEffect}
\end{figure}

Terrain aspect, jointly measured by the \emph{Eastness} and \emph{Northness} covariates, both expressed by their mean and standard deviations, played a significant role albeit with a small amplitude (\ie, small absolute coefficient values). According to their sign, SUs facing North or West were related to larger landslide intensities and larger susceptibility estimates. We note here that the decomposition of the terrain aspect into its two main components (\emph{Eastness} and \emph{Northness}) was a numerically convenient way to handle the nonlinear and cyclic exposition signal on slope stability/instability conditions. However, by decoupling the aspect into a linear combination of sine and cosine components, we lost the original interpretation of the overall effect of the aspect expressed in degrees within the $[0, 360)^\circ$ range. To compensate for this, in Figure~\ref{fig:AspectEffect} we show the overall reconstructed effect of the terrain aspect, for each model. Precisely, in the Figure~we plot $\beta_{\text{Eastness}}\cos(\theta)+\beta_{\text{Northness}}\sin(\theta)$ as a function of  $\theta\in[0,360]^\circ$, where $\beta_{\text{Eastness}}$ and $\beta_{\text{Northness}}$ denote 
the \emph{Eastness} and \emph{Northness} coefficients, respectively, estimated from each model.
Inspection of the Figure~reveals that the effect of terrain aspect on landslide intensity and susceptibility is significant, and when back-transformed to its original scale, the W-NW components mentioned above reveal a clear positive contribution to the landslide counts (\ie, landslide intensity), which changes to a negative effect when moving towards E-SE components. This was known in the study area, and depends on the geometric and geomorphological interaction between the prevalent attitude of the bedding planes that characterise the study area and the orientation and geometry of the slopes \citep{marchesini2015assessing,santangelo-MethodAssessment-2015}.

\subsection{Predictive performance of models} 
\label{sec:CV}

The goodness-of-fit for the baseline models Mod1 and Mod2 was weak in terms of fitted counts, but acceptable in terms of binary metrics. Conversely, the more advanced random effect models Mod3, Mod4, and Mod5 showed outstanding explanatory performances both in terms of expected landslide counts, and fitted presence-absence probabilities (see Figure~\ref{fig:CountFit}). Thus, it is natural to wonder whether or not the more complex LGCP models overfit the data. In case of overfitting, their predictive performance would be low.

To assess this, and to quantify the predictive performance of each model, we designed two cross-validation (CV) procedures. Because the data are spatio-temporal in nature, we considered both a spatial CV scheme and a temporal CV scheme. We measured the spatial predictive performance using a 10-fold cross-validation procedure (abbreviated \textit{Space 10-Fold}), and the temporal predictive performance using a leave-one-out cross-validation procedure (abbreviated \textit{Time leave-one-out}). More precisely, the spatial 10-fold CV consists in splitting the original dataset into 10 complementary subsets at random, each comprising 10\% of the original SUs. The model is then fitted using nine subsets (\ie, 90\% of the SUs), and subsequently validated using the left-out subset (\ie, 10\% of the SUs). The procedure is repeated by leaving out each subset. The complementary constraint ensures that every SU in the Collazzone study area is predicted exactly once (and only once) for all time intervals during the CV routine. As for the temporal leave-one-out CV, we leave out the data from one of the six time intervals, then we fit the model using the five remaining intervals, and finally we validate the model using the data from the time interval that was left out. The procedure is repeated by leaving out the data from each time interval. Essentially, this corresponds to a temporal 6-fold CV scheme, where each test set consists of a single time interval.

\begin{figure}[!htbp]
	\centering
	\includegraphics[width=0.9\linewidth]{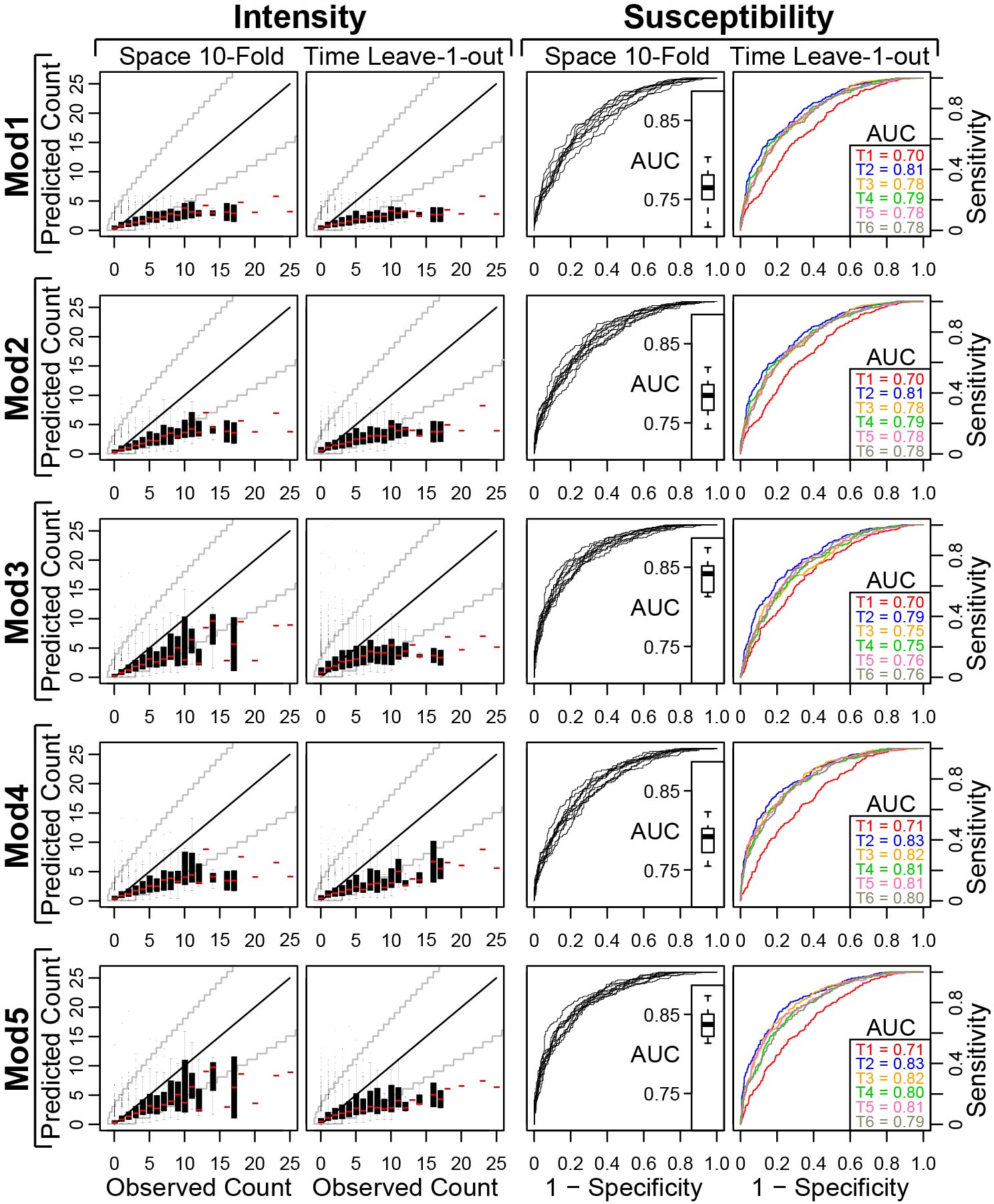}
	\caption{Out-of-sample performance summary of each fitted LGCP model Mod1 to Mod5 (top to bottom). Predicted counts plotted against observed counts based on a spatial 10-fold CV scheme (Column 1) and a temporal leave-one-out CV scheme (Column 2). For each observed count, we show boxplots of predicted counts to summarise their distribution across all SUs. Grey lines are 95\% intervals for an exact Poisson distribution with mean between 0 and 25. ROC curves and corresponding AUC values for each test set based on a spatial 10-fold CV scheme (Column 3) and a temporal leave-one-out CV scheme (Column 4).} 
	\label{fig:CVs}
\end{figure}

For both CV schemes, we examined the models' performances both in terms of intensity (\ie, predicted counts) and susceptibility (\ie, predicted probability of landslide occurrence), using the same summary measures used in Figure~\ref{fig:CountFit}. Results are summarised in Figure~\ref{fig:CVs} where the two main vertical panels represent the intensity and susceptibility results, and the sub-columns summarise the results for the \textit{Space 10-fold} and \textit{Time leave-one-out} procedures. The different rows correspond to Mod1, Mod2, Mod3, Mod4, and Mod5 (from top to bottom). We opted to avoid colour coding the \textit{Time leave-one-out} intensities by time to improve readability of the figure. 

Inspection of Figure~\ref{fig:CVs} reveals that the agreement between the observed and the predicted landslide counts increases significantly from our simple models (Mod1, Mod2) to the advanced models that include latent variables (Mod3, Mod4, Mod5), which are better capable of predicting the number of landslides in each SU. In Mod3 and Mod5, the match between observed and estimated counts is reasonably good, \ie, it is still confined within the theoretical uncertainty of the Poisson distribution, up to $14$ landslides. Conversely, from this number to the maximum ($25$ landslides) in the dataset, the predicted counts tend to underestimate the actual observations. This is not a feature of the model, but rather a consequence from our dataset, which comprises a large number of SUs with no or a few landslides and very few SUs with many landslides. Inspection of Figure~\ref{fig:AggrByTime} reveals that the landslide dataset has only a few isolated samples larger than $10$ counts, which is where the model starts to perform poorly. Also, the \textit{Time leave-one-out} is the CV scheme that deviates most strongly from the performance obtained for the fit. An explanation is that the \textit{Space 10-fold} removes 533 SUs per iteration ($889\times6/10 \approx 533$), whereas a temporal CV removes $889$ SUs. Moreover, the temporal CV disrupts the coherence in the data more strongly, since removing one time interval removes either $50\%$ or $100\%$ of the direct temporal neighbours for a large number of data points. Therefore, it is the most challenging CV scenario we could devise. Nevertheless, we note that the overall performance shown in the susceptibility case still falls in the ``excellent" class of \citet{hosmer2000}, with outstanding AUC metrics. The improvement from Mod1 to Mod5 is measured quantitatively by the models' AUCs, which justify the inclusion of the latent effects. 

\subsection{Best intensity--susceptibility predictive model---Mod5} 
\label{sec:best-model}

Examining the results or our modelling effort in terms of \i estimated landslide counts, \ie, of predicted landslide intensity \citep{Lombardo.etal:2018,lombardo2019numerical,lombardo2019geostatistical}, and of \ii estimated binary presence-absence of landslides, \ie, of predicted landslide susceptibility \citep{brabb1985,guzzetti1999landslidehazard,reichenbach2018}, and considering the models' spatial and temporal structures (Figures~\ref{fig:Fig5}, \ref{fig:Fig6}), and their fitting (Figure~\ref{fig:CountFit}) and predicting (Figure~\ref{fig:CVs}) performances, we maintain that Mod5---which jointly accounts for the spatial and temporal dependencies---has a better overall performance than the other four models (Mod1 to Mod4). 

\begin{figure}[!htbp]
	\centering
	\includegraphics[width=0.55\linewidth]{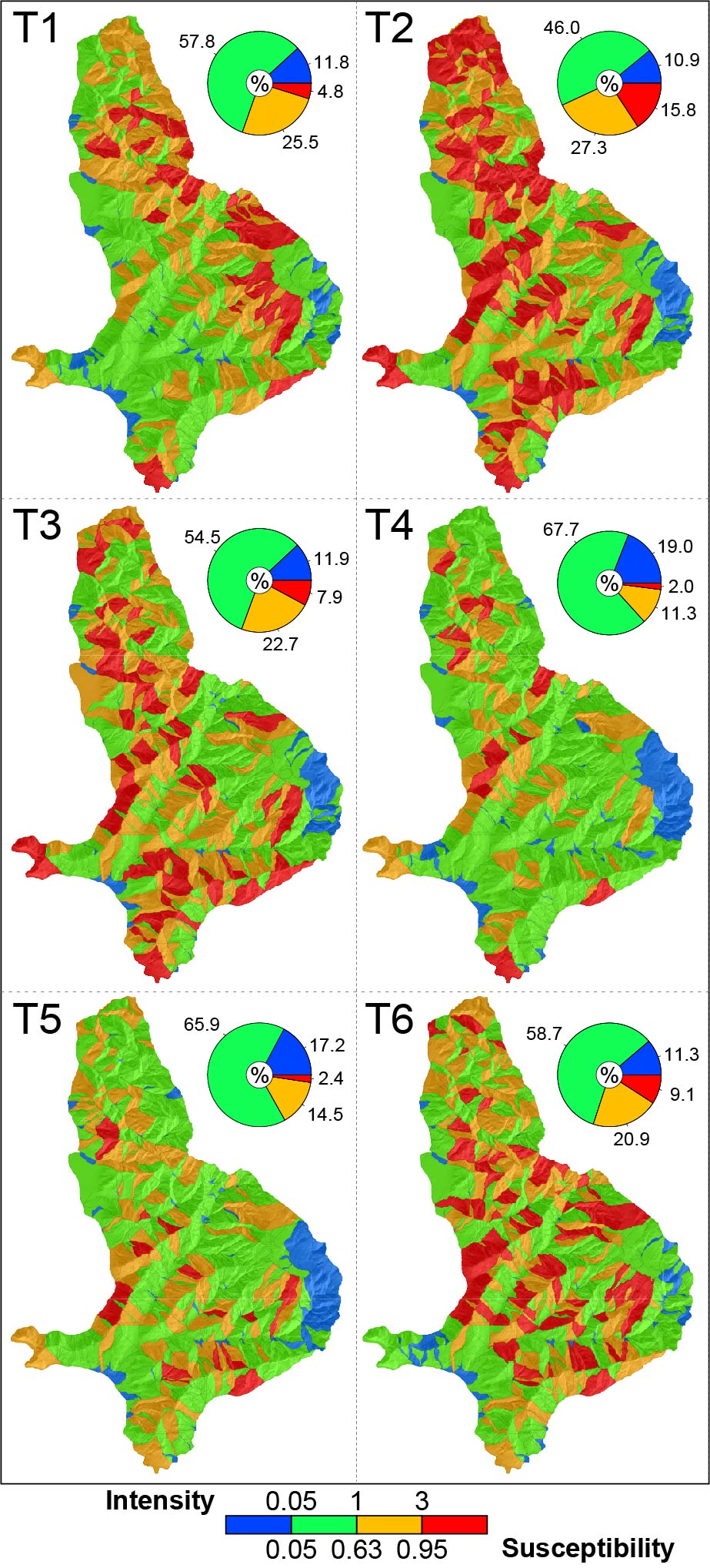}
	\caption{Combined intensity--susceptibility classification of the Collazzone study area, Umbria, Central Italy, based on model Mod5 constructed using morphometric, geologic, bedding attitude, and space-time dependencies. 
	Each map corresponds to one of the six time periods shown in Figure~\ref{fig:AggrByTime}.
	Pie-charts show the percentage of SUs falling into one of the four considered classes. 
	See \S\ref{sec:Classification} for an explanation of the adopted classification and ranking scheme.}
	\label{fig:SummaryPredMaps}
\end{figure}

Mod5 provides comparable patterns to Mod3 in terms of the predicted landslide counts over time, keeping the flexibility of Mod4 in the binary (\ie, susceptibility) predictions. We conclude that Mod5 is our best model, and we select it to generate a combined landslide intensity--susceptibility classification for our study area, adopting the ranking scheme proposed in \S\ref{sec:Classification}. We summarise the results of Mod5, for each of the six time periods (T1--T6), in Table~\ref{tab:SummaryClassTab} and Figure~\ref{fig:SummaryPredMaps}. Inspection of results reveals some degree of temporal variability in the combined intensity--susceptibility patterns. However, the general spatio-temporal pattern remains about the same. Similarly to the outcomes of our baseline model Mod1 (\S\ref{sec:BaselineResults}), the number of SUs that Mod5 predicts capable of generating a large or very large number of landslides is limited, whereas the number of SUs that can generate landslides, \ie, that are potentially ``susceptible" to slope failures, is large and geographically distributed. This is reasonable from a geomorphological and landscape evolution perspectives. 

Overall, and for the entire considered period (Figure~\ref{fig:AggrByTime}), model Mod5 classifies the (relative or absolute) majority of the SUs, from 409 (T2, 46.0\%) to 602 (T4, 67.5\%), as of ``Uncertain Type 1" (\S\ref{sec:Classification}). In each of these SUs, covering collectively between 25.6 (32.4\%) and 49.8 (63.1\%) $km^2$, the estimated landslide intensity, \ie, the expected number of landslides, is in the range $(0.05,1]$, on average. The second class encompasses from 100 (T4, 11.3\%) to 227 (T1, 25.5\%) SUs classified as of ``Uncertain Type 2". In these SUs, covering collectively between 17.8 and 25.5 $km^2$ (22.5 to 32.3\%), the number of expected landslides is in the range $(1,3]$, on average (Table~\ref{tab:SummaryClassTab}). With a few exceptions, model Mod5 classifies the smallest number of SUs, from only 18 (T4, 2.0\%) to 140 (T6, 15.8\%), as  ``Clearly Unstable", covering between 4.6 (5.8\%) and 23.6 (29.9\%) $km^2$, followed by from 97 (T2, 10.9\%) to 169 (T4, 19.0\%) SUs classified as  ``Clearly Stable", covering between 3.5 (4.4\%) and 6.8 (8.7\%) $km^2$. We maintain that the variability in the predicted classification estimates (Figure~\ref{fig:SummaryPredMaps}, Table~\ref{tab:SummaryClassTab}) measures \i the spatio-temporal uncertainty of the landslide intensity--susceptibility estimates at the spatial scale and in the temporal range considered by our modelling experiment, \ie, the aleatory uncertainty inherent in the landslide processes, and \ii the epistemic model uncertainty introduced by the adopted modelling framework and the available landslide (Figure~\ref{fig:StudyArea}) and thematic data (Table~\ref{tab:covariates}).

\begin{table}[!htbp]
\centering
\caption{Summary of the unified intensity--susceptibility classes, from T1 to T6. See Figure~\ref{fig:AggrByTime} for coverage of time periods.}
\label{tab:SummaryClassTab}
\begin{tabular}{llcccc}
& & & & & \\ \hline
Time period & Class & Slope units (SUs) & & Total SU area & \\
& & Count (\#) & \% & $km^2$ & \% \\ \hline
\vspace{-0.4cm}
& & & & &\\
T1 & Clearly stable & 105 & 11.8 & 3.0 & 3.8 \\
before 1941 & Uncertain Type 1 & 514 & 57.8 & 40.3 & 51.1 \\ 
& Uncertain Type 2 & 227 & 25.5 & 25.5 & 32.3 \\ 
& Clearly unstable & 43 & 4.8 & 10.1 & 12.8 \\
\vspace{-0.253cm}
& & & & &\\ \hline
T2 & Clearly stable & 97 & 10.9 & 3.5 & 4.4 \\
1941-1954 & Uncertain Type 1 & 409 & 46.0 & 25.6 & 32.4 \\ 
 & Uncertain Type 2 & 243 & 27.3 & 26.3 & 33.3 \\ 
 & Clearly unstable & 140 & 15.8  & 23.6 & 29.9 \\
\vspace{-0.3cm} 
& & & & &\\ \hline
T3 & Clearly stable & 106 & 11.9  & 3.6 & 4.6 \\
1954-1977 & Uncertain Type 1 & 511 & 57.5 & 31.7 & 40.1 \\ 
 & Uncertain Type 2 & 202 & 22.7 & 26.1 & 33.0 \\ 
 & Clearly unstable & 70 & 7.9 & 17.6 & 22.3 \\
\vspace{-0.3cm}
& & & & &\\ \hline
T4 & Clearly stable & 169 & 19.0 & 6.8 & 8.7 \\
 1977-1996& Uncertain Type 1 & 602 & 67.7  & 49.8 & 63.1 \\ 
 & Uncertain Type 2 & 100 & 11.3  & 17.8 & 22.5 \\ 
 & Clearly unstable & 18 & 2.0  & 4.6 & 5.8 \\
\vspace{-0.3cm}
& & & & &\\ \hline
T5 & Clearly stable & 153 & 17.2  & 5.7 & 7.2 \\
 1997 (snow) & Uncertain Type 1 & 586 & 65.9  & 45.4 & 57.6 \\ 
 & Uncertain Type 2 & 129 & 14.5  & 23.5 & 29.8 \\ 
 & Clearly unstable & 21 & 2.4 & 4.3 & 5.4 \\
\vspace{-0.3cm}
& & & & &\\ \hline
T6 & Clearly stable & 100 & 11.3  & 2.6 & 3.3 \\
1998-2014 & Uncertain Type 1 & 522 & 58.7  & 37.2 & 47.2 \\ 
 & Uncertain Type 2 & 186 & 20.9  & 22.2 & 28.1 \\ 
 & Clearly unstable & 81 & 9.1 & 16.9 & 21.4 \\ \hline
\end{tabular}
\end{table}

\subsection{Computational requirements}
\label{sec:CR}

All models presented in this paper fitted on any state-of-the-art computer. Running times were less than one minute (baseline models Mod1, Mod2), several minutes (Mod4), and around 6 hours (Mod3, Mod5). For estimation of the full model with all data and for the cross-validation models with some of the landslide counts held out, running times were of comparable order of magnitude in all cases. In particular, we consistently observed long running times of several hours for Mod3 and Mod5 due to longer computations related to Laplace approximations, but the diagnostic output of \texttt{R-INLA} did not indicate any instabilities. Memory requirements were less than 1Gb in all cases when using 2 cores in parallel for each job. 

The main computational cost of our modelling procedure stems from the Laplace approximations, performed repeatedly during an estimation run with INLA. A general rule of thumb is that computation times increase with the number of observations (5334 with our dataset), the number of latent variables (29 parameters for Mod1, 34 parameters for Mod2, and 5368 parameters in the Mod3, Mod4 and Mod5 comprising random effects), and the complexity of the additive predictor model comprising components for fixed and random effects. When there is a tendency towards confounding of effects, \ie, when different additive components can provide similar contributions to the predictor, computations can become less stable such that computation times increase, or estimation may even fail, which did not happen with our models. 

The mapping units for which event counts are recorded may have higher resolution or may span larger areas than here, with up to several hundreds of thousands of observations \citep[\eg,][]{Lombardo.etal:2018}. Running INLA on such datasets, with models comprising several thousands of latent variables, then typically takes several hours, or even several days in extreme cases. With memory requirements of \texttt{R-INLA} easily exceeding 16Gb in such high-dimensional models, estimation is usually carried out on machines dedicated to scientific computing. In general, Bayesian hierarchical modelling demands considerably higher computing resources than classical frequentist approaches but also provides significant benefits as shown in this work. In the context of geomorphological applications, the estimation results usually have to be established only once, without any need to reestimate models for continuously updated data, such that this increased computational burden remains manageable in practice. Moreover, the library \texttt{R-INLA} provides several choices of less accurate approximation schemes to speed up estimation and reduce memory usage. Its recent integration of the powerful \texttt{PARDISO} library for numerical matrix computations further increases its potential for solving very high-dimensional problems \citep{Niekerk.etsal.2019}. 	
	
\section{Discussion}
\label{sec:Discussion}

We now discuss the results of our modelling experiment. We first focus on what we consider the main advantages and the limitations of the new modelling framework, and its potential applicability to other areas (\S\ref{sec:DiscussionModelling}). Next, we make specific and general considerations on the results of our work for landslide hazard assessments (\S\ref{sec:DiscussionHazard}). This is followed by a critical analysis of the modelling approach for geomorphological and slope stability inference (\S\ref{sec:DiscussionGeomorphology}). Lastly, we provide a perspective on further developments of landslide predictive modelling and zoning (\S\ref{sec:DiscussionPerspective}).  

\subsection{A new landslide predictive modelling framework}
\label{sec:DiscussionModelling}

In our work, we experimented an innovative, Bayesian modelling framework for the spatio-temporal prediction of landslides of the slide type \citep{Hungr2014} (\S\ref{sec:LGCP}). Results proved that the adopted framework was capable of predicting the temporal, the spatial, and the spatio-temporal distributions of known landslides that occurred in our study area in the period from before 1941 to 2014 (Figure~\ref{fig:StudyArea}). Results also showed that considering the existing, albeit not identified explicitly (\ie, ``latent"), landslide dependencies in space and time improved significantly the model predictive performance (\S\ref{sec:CV} and Figure~\ref{fig:CVs}), when compared to a simpler (``traditional") model, exemplified by our baseline model Mod1 (Figure~\ref{fig:BaselineMap}) which does not consider the spatio-temporal dependencies among landslides. 

Our Log-Gaussian Cox Process (LGCP) model is a ``doubly" stochastic process, with the stochasticity given by \i a Poisson component, describing the number of landslides in each SU, and by \ii a Gaussian component, which describes the landslide intensity (\ie, the expected count per SU) on the logarithmic scale and allows to incorporate several types of random effects, namely linear, nonlinear and nonlinear, at a latent level \citep{Lombardo.etal:2018}. Therefore, the LGCP approach provides a convenient framework to investigate and interpret 
morphometric and thematic covariates' influence on slope instability in the study area (Figure~\ref{fig:Coefficients}). It also brings to light unobserved dependencies that influence the landslide intensity function $\Lambda_j(s)$ (\S\ref{sec:BayesianModelling}), and the derived landslide susceptibility estimates. We consider this a significant improvement over existing, statistically-based landslide prediction modelling tools currently available in the literature \citep{reichenbach2018}, which do not cope with latent effects, and typically do not consider complex temporal or spatio-temporal landslide dependencies. 

To construct our models, we exploited a multi-temporal landslide inventory comprised of numerous (3,379) landslides, which occurred in a significantly long period over our study area (Figures \ref{fig:StudyArea}, \ref{fig:AggrByTime}). The geographical (``cartographic") and thematic (``geomorphological") detail and the accuracy of the landslide mapping were key to inform properly our models, and to evaluate their performances (Figure~\ref{fig:CountFit}) and prediction skills (Figure~\ref{fig:CVs}). This may be seen as a limitation of the proposed framework, which requires accurate landslide data to provide reliable intensity and susceptibility estimates. However, we maintain that in order to predict landslides in space and time such detailed and accurate information is necessary (mandatory), albeit it may be costly and time consuming to obtain it \citep{galli-ComparingLandslide-2008,guzzetti2012}. While accurate multi-temporal landslide information is available, together with relevant thematic data (\eg, Table~\ref{tab:covariates}), the added effort to construct and run an advanced model (\eg, our model Mod5) compared to a simpler model (\eg, Mod1) is negligible, both in terms of GIS pre-processing and data preparation, and for the statistical modelling. Indeed, with the \texttt{R-INLA} library, we can run very complex models using a simple syntax. In our case, the difference between Mod1 and Mod5 was two additional lines of code.  

Inspection of the model fitting (Figure~\ref{fig:CountFit}) and predictive (Figure~\ref{fig:CVs}) performances further reveals that the more advanced models (Mod3, Mod4, Mod5) where generally better at predicting the spatial (\ie, ``where") rather than the temporal (\ie, ``when") component. We maintain that this is due to the combined effect of \i the inherent short-term temporal viability---and related unpredictability---of landslide phenomena at the SU scale, at least in our study area \citep{samia-CharacterizationQuantification-2017,samia-LandslidesFollow-2017,Samia2018}, and \ii the number and length of the considered temporal periods and the number of landslides in each period (Figure~\ref{fig:AggrByTime}), which depend on the temporal frequency of landslides in our study area. The latter, is in turn controlled by the frequency of the landslide triggering forcing events (\eg, severe or prolonged rainfall periods, rapid snow-melt events) \citep{rossi2010,witt2010}. This has hazard and geomorphological consequences, which we address below in \S\ref{sec:DiscussionHazard} and \S\ref{sec:DiscussionGeomorphology}.

The models accounting of the spatial landslide dependencies involve smoothing residuals across adjacent SUs. This may have introduced some local inconsistency or error at the boundary between the SUs. However, as mentioned in \S\ref{sec:Space}, the models are flexible, and let the data prevail without introducing too much noise to the intensity estimates. Should one need to account for the effect of a ``rigid" barrier (\eg, a river, a major divide, a main lithologic or tectonic discontinuity) on landslide intensity or susceptibility, two solutions are possible. \citet{bakka2019} have developed a model for incorporating physical barriers, which can be fitted using \texttt{R-INLA}, although their method relies on a different type of spatial effect as the one exploited in this work. Alternatively, one can remove manually the links between adjacent SUs in the adjacency matrix (Figure~\ref{fig:AdjMat}). Since the residual smoothing process is governed by the adjacency matrix, removing appropriate links will prevent the latent effect from ``propagating'' from one SU to its direct neighbors. 

The computational burden of a space-time LGCP model in \texttt{R-INLA} depends on, and scales with, the size of the dataset by exploiting random effects with sparse precision (\ie, inverse covariance) matrices. Relatively small areas like the one used for our experiment can be investigated effectively with a standard, modern personal computer even for spatio-temporal models (\eg, model Mod5). Larger datasets covering large and very large areas need proportionally larger computer facilities. We note here that the adoption of the SUs as the mapping unit of reference, or of other similar terrain mapping units \citep{guzzetti1999landslidehazard,guzzetti2005,van2006}, as opposed to ``grid cells" or pixels, has reduced greatly the computational burden. In general, use of SUs facilitates the construction of complex models even for large and very large areas covering thousands of square kilometres \citep{alvioli2016automatic}. We conclude that the main limitation for performing complex, space-time LGCP landslide modelling is mostly due to the lack of accurate datasets and detailed multi-temporal landslide inventories, and not to the computational requirements. This should guide geomorphologists interested in landslide prediction modelling in their time allocation and resource investment \citep{guzzetti1999landslidehazard}. 

We further note that we successfully tested the new framework (Figure~\ref{fig:CVs}) for landslides predominantly of the ``slide" type \citep{Hungr2014}, which are common and abundant in our study area, and in similar areas in Central Italy and elsewhere in similar physiographical settings. We acknowledge that further efforts are required to test the framework with different landslide types, since their temporal and spatial dependencies may vary. However, we do not see any geomorphological or statistical reason that should limit or hamper the applicability of the proposed framework to other landslide types. Lastly, we stress that the predictions made by all our models are valid under the general assumption that \i the driving forces that control the landslide processes in the study area are known and captured through the covariates used in the models, and \ii the driving forces will remain nearly the same in the foreseeable future \citep{fabbri-PredictionFuture-2003,Guzzetti2006a}. We maintain that both assumptions hold in our study area, but the same key assumptions should be considered thoroughly when similar models are constructed in other areas. 

\subsection{Statistical considerations}
\label{sec:StatConsideration}

A rigorous implementation of a model based on spatial point pattern theory would have required to treat each landslide as a precisely geolocated point. For practical implementation, it is usually satisfactory to know in which mapping unit a landslide occurred. 
However, a few landslides in the multi-temporal inventory (Figure~\ref{fig:StudyArea}), mostly present in the T1 and T2 periods (Figure~\ref{fig:AggrByTime}), have a large or very large area (for T1 and T2: $A_L$ $\geq$  $2.2\times10^2$ $m^2$; for T3, T4 and T5: $A_L$ $\geq$  $5.8\times10^2$ $m^2$), and intersect multiple SUs. Treating such large landslides as single ``points" would have been a severe forcing from a geomorphological perspective.
We were then faced with the choice of whether to conflict \i with the conditional independence assumption of points in our modelling tool, or \ii  with the empirical, geomorphological field evidence. 

The conditional independence assumption states that observed landslide counts are independent if we know the value of the predictor comprising the covariate information and random effects (the latter only if they are part of the model). This assumption is common to all well-established spatial statistical models for discrete data, irrespective of the choice of a susceptility model (\ie, using a Bernoulli distribution for presence-absence data) or an intensity model (\ie, using a Poisson distribution for count data), and seems difficult to abandon, especially as it is a critical assumption for using INLA. 

By analogy with susceptibility studies in the existing literature \citep[e.g.,][]{Guzzetti2006a}, we chose to respect the field evidence, and we counted the presence of a landslide---or of a portion of a landslide---in each SU if the landslide area exceeded 2\% of the SU area, a percentage that accounts for possible mapping errors \citep{Carrara1991,carrara1995gis}. We acknowledge that this approach creates some dependence between events in nearby SUs and can lead to local clustering patterns of events that cannot be fully captured by models that obey the conditional independence assumption. This entails two major problems: first, the model lacks realistic small-scale behavior and may underestimate components of landslide risk at relatively small spatial scales; second, statistical inference is flawed by considering dependent observations as independent, which will cause an underestimation of uncertainty, for instance by declaring certain covariates as significant while they are not.

However, by using random effects as in our models we can substantially alleviate these problems by capturing local dependence and clustering structures that cannot be explained by geomorphological covariates alone. In other words, the latent spatial random effect can capture (part of) the dependence induced by the largest landslides affecting several SUs. While classical generalized linear models (GLMs) have only fixed effects and assume complete independence of observations, our models are based on the less restrictive assumption of conditional independence with respect to the combination of fixed and random effects. In this Bayesian framework, we can model the propagation of landslide counts over neighboring SUs, which also includes how far a single 
landslide extends in space (\ie, ``how large'' it is). By working with intensities instead of susceptibilities, \ie, with count data instead of presence-absence data, we further reduce the loss of information in data and models when opting for larger mapping units, where the phenomenon of single landslides stretching over several units becomes less frequent. 
As a result, the procedure of using an intensity framework with random effects brings our models closer to the accepted definition of landslide hazard \citep{varnes1984,guzzetti1999landslidehazard, guzzetti2005probabilistic}.

\subsection{Hazard considerations}
\label{sec:DiscussionHazard}

As mentioned in \S\ref{sec:landslide-prediction}, the prediction of landslide hazard proposed by \citet{varnes1984}, and later modified by \citet{guzzetti1999landslidehazard} and \citet{guzzetti2005probabilistic}, requires the anticipation---in probabilistic terms---of ``where" (spatial component), ``when" (temporal component), and ``how large" or destructive (magnitude component) landslides are expected to be in an area \citep{guzzetti2005}. Our new modelling framework, and specifically our more advanced model Mod5, fulfils the definition, to a large extent. Model Mod5 accounts for the spatial and temporal dependencies of landslides, including latent effects not explicitly described by other covariates. The magnitude component of the hazard is also present in model Mod5, given by the expected number landslides in each SU, \ie, by the landslide intensity. We note here that the number of landslides was used as a measure of landslide event magnitude, \eg, by \citet{keefer1984} for earthquake-induced landslides and by \citet{malamud2004} for landslides caused by weather and geophysical triggers. Furthermore, landslide size characteristics, including landslide area \citep{hovius1997,guzzetti-PowerlawCorrelations-2002,malamud2004}, volume \citep{malamud2004,brunetti2009}, area-to-volume ratio \citep{guzzetti2009,larsen2010}, and length-to-width ratio \citep{taylor2018}, which can all be associated to the vulnerability to landslides \citep{galli-LandslideVulnerability-2007} and hence to the landslide destructive power, are known to be empirically related to the number of landslides in an area. And, because we also considered the landslides' extent when counting slope instabilities per slope units, our models Mod3 and Mod5 are even informed by the latent fields on the persistence (a proxy for size) of landslide counts over space.

We conclude that the landslide intensity framework proposed by \citet{Lombardo.etal:2018,lombardo2019numerical,lombardo2019geostatistical} for spatial predictions, and extended here in time and space-time, is well suited to fulfil the requirements given by the standard definition of landslide hazard, and capable to do so within a single model. This is a significant advancement over previous hazard models that considered the spatial, temporal, and the landslide area (a proxy for magnitude) components separately \citep{guzzetti2005probabilistic,Guzzetti2006a}, and had to further assume the independence of the three components to properly estimate landslide hazard in probabilistic terms. 

Our experiment revealed that only a few SUs in our study area exhibited a constant landslide clustering or repellency trend over the entire considered period, and that most of the SUs exhibited a (randomly) varying clustering/repellency signal (Figure~\ref{fig:TimeDependence}). We take this as empirical evidence of the fact that the temporal prediction of landslides over relatively long periods, longer than about 15 years in our case, is problematic and inherently uncertain. \citet{samia-CharacterizationQuantification-2017,samia-LandslidesFollow-2017,Samia2018}, working in the same study area, identified a landslide heritage effect---which they called ``landslide path dependency"---that conditions the occurrence of new landslides dependent on the location of previous landslides in the same SU, over periods of less than 15 years. Both empirical findings have consequences for hazard assessment. Neglecting the temporal dependence on landslides will underestimate hazard in SUs characterised by a clustering effect (inflating ``type II" errors), and will overestimate  hazard in SUs characterised by a repellent effect (inflating ``type I" errors). We further note that common approaches to predict future landslide occurrences over large areas, including the definition of empirical landslide thresholds for the possible initiation of landslides from landslide and rainfall records \citep{aleotti2004warning,guzzetti2007,guzzetti2008,saito2010,ko2018,segoni-ReviewRecent-2018}, and the calculation of return periods from time-series of triggering events, chiefly rainfall or precipitation events \citep{frattini2009}, assume the stationarity of the landslide processes over time. However, evidence shows that landslide processes are not stationary in our study area, and arguably in other similar areas in Central Italy and in other similar physiographical and climatic settings. The finding poses questions on the reliability of landslide forecast and prediction models based on past landslide and rainfall records \citep{rossi2010,witt2010,segoni-ReviewRecent-2018,guzzetti-LEWS-2019}. 

Lastly, we note that our work introduced two novel advancements in landslide hazard modelling. First, we provided a robust way to classify landslide susceptibility, obtained from the landslide intensity using equation \eqref{eq:conversion}. The approach avoids the common problem of interpreting intermediate probability estimates as a measure of intermediate or ``mean" or ``moderate" susceptibility levels, which is incorrect, conceptually and operationally, and can lead to serious problems if the susceptibility models and associated zonings are used for practical applications \citep{guzzetti-ComparingLandslide-2000,galli-ComparingLandslide-2008,reichenbach2018}. Second, we propose a new way of portraying in a single map the information provided jointly by the landslide intensity and the landslide susceptibility estimates. We maintain that the use of a single cartographic representation to show the combined intensity--susceptibility information facilitates the use of the zonation for practical applications, and the design of landslide protocols for land planning and management \citep{guzzetti-ComparingLandslide-2000,reichenbach2018}. To the best of our knowledge, the combined intensity--susceptibility classification proposed in \S\ref{sec:Classification} and exemplified in Figure~\ref{fig:SummaryPredMaps} for our best model Mod5, are unique in the landslide hazard modelling literature.

\subsection{Geomorphological considerations}
\label{sec:DiscussionGeomorphology}

The performance of statistically-based landslide prediction models depend entirely on the models structure and on the data used to inform them. If the data (the ``covariates") are accurate and meaningful, an analysis of the model results can provide valuable insights on the geomorphic processes that control the spatio-temporal distribution of landslides in area, provided the modelling framework is geomorphologically sound. 

Concerning data, to inform our models we used covariates that are known to represent geomorphic conditions that favour or hamper the formation of landslides in our study area \citep{Guzzetti2006a,Guzzetti2006b,Ardizzone2007,galli-ComparingLandslide-2008}, in similar geologic, physiographic, and climatic settings \citep{Carrara1991,carrara2003,carro2003,guzzetti2005,marchesini2014}, and even in very different landscapes  \citep{budimir-SystematicReview-2015,goetz2015evaluating,lombardo2016b,reichenbach2018}. With this respect, we maintain that our morphometric, lithological, and structural covariates (Table~\ref{tab:covariates}) are sound, accurate, and meaningful landslide predictors, and that they contribute to explain the known spatio-temporal distribution of landslides in our (Figure~\ref{fig:StudyArea}) and in similar study areas.

Concerning the model structure, the LGCP framework assumes that individual landslides in a complex landscape are the result of a point process, in space and time. In this framework, a single landslide, \ie, a single element of a large population of landslides, is represented by a ``point" $(s_i,t_i)$ defined by its spatial ($s_i$) and temporal ($t_i$) location, \ie, ``where" and ``when" the ``point" landslide occurred in the investigated area (Figure~\ref{fig:StudyArea}) and period (Figure~\ref{fig:AggrByTime}). The model further assumes that the spatio-temporal distribution of landslide points is the result of an unobserved intensity function ($\lambda(s,t)$) that varies over space and time. It is the stochastic variation of this intensity function that determined the location and temporal occurrence of the landslides. The last assumption is that the model incorporates effects carried directly by the data, \ie, by the covariates, and by unobserved random effects not explained directly by the covariates. In our case, such random effects include, \eg, the fact that geomorphologically similar and adjacent SUs behave similarly in their ability to generate landslides, and the fact that landslides tend to repeat in time in the same places where they occurred in the past \citep{Samia2018}. Overall, these modelling assumptions are reasonable, from a geomorphological perspective. 

Most of the landslides in the Collazzone study area have an area smaller than $Mo=5,648 m^2$, about 0.01\% of the size of the study area. Even the largest landslide, extending for $1.5\times10^6$ $m^2$, covers less than 2\% of the study area. In the study area landslides are caused chiefly by severe weather events, each covering a small or very small fraction of a year, and hence an even much smaller fraction of the multi-decadal period considered by our analyses. We conclude that for the spatial and temporal evolution of the landscape that characterises the Collazzone study area, individual landslides are---or can be safely considered as---``point" events, both in space and time. 

It is known that landslides in the study area are not distributed randomly in space (Figure~\ref{fig:StudyArea}) \citep{Guzzetti2006a,Guzzetti2006b,Ardizzone2007,galli-ComparingLandslide-2008}, and that the size and type of the landslides are controlled by the interaction between the geometry of the slopes (chiefly terrain gradient and aspect) and the attitude of the main lithological layers (\ie, the strike and dip of sand and gravel levels, and clay laminations) \citep{Guzzetti2006a,marchesini2015assessing,santangelo-MethodAssessment-2015}. Thus, adjacent ``anaclinal" slopes tend to generate similar, large, deep-seated slides, or complex and compound landslides, whereas adjacent ``cataclinal" slopes tend to generate similar, small shallow slides and minor rotational landslides.
It is also known that landslides in the area do not occur randomly in time. As mentioned before, \citet{samia-CharacterizationQuantification-2017,samia-LandslidesFollow-2017,Samia2018}, who worked in the same area, identified a landslide heritage effect that conditions the occurrence of new landslides dependent on the location of previous landslides over periods of less than 15 years.
Our own results confirm that this heritage effect is limited in time, with only a minority of the SUs exhibiting a constant, long term clustering or repellency trend, with  the vast majority of the SUs showing fluctuating dependence signals through time (Figure~\ref{fig:TimeDependence}). Indeed, this is a reasonable and expected behaviour for the medium to long-term evolution of  slopes, and more generally of landscapes shaped by mass wasting processes. We conclude that the assumption that there exist ``unobserved" latent effects that control and explain the spatio-temporal distribution of landslides is geomorphologically sound, and it matches and explains the existing empirical evidences.

We see two main limitations of geomorphological relevance of our current LGCP framework. First, we do not explicitly consider the size (\eg, area, volume) of the predicted landslides in each SU, with consequences on the possibility to exploit the modelling results for erosion, sediment and landscape evolution modelling. Second, the applicability of the model over very long periods (centuries or millennia) remains to be determined. For the former, new modelling frameworks will have to be devised, and tested. For the latter, not only we lack long-term past landslide data to train sound models, but we also lack a proper understanding of how climate may change and influence future slope instabilities, in the same general area \citep{alvioli-ImplicationsClimate-2018}, and in other areas \citep{gariano-LandslidesChanging-2016}. The main problem to overcome both limitations lays in the lack of accurate, spatially distributed, multi-temporal landslide datasets. However, the rapidly improving methods and techniques for the automatic or semi-automatic detection and mapping of landslides over large areas from remotely sensed data promise to bridge this data acquisition gap \citep{guzzetti2012}.  

\subsection{Perspective}
\label{sec:DiscussionPerspective}

We see a number of possible future improvements to our work, with further specific and general modelling, hazard, and geomorphological implications. 
For the specific case of the Collazzone study area, we envision adding new covariates to the model, including covariates describing \i land use and land cover types, which are known to influence the size, abundance, and frequency of slope failures in the study area, and \ii the morphometric and hydrological settings of the individual SUs, which can also influence the presence and evolution of landslides in layered sediments \citep{Carrara1991}.
An additional improvement will be to add covariates describing spatio-temporal environmental variations, including, \eg, space-time changes in land use and land cover driven by different agricultural or forest practices. 
We also envision improving our modelling of the spatial latent effect introduced by SUs with similar or different  lithological, hydrological, or structural characteristics. For the purpose, we could experiment the incorporation of physical barriers (\eg, lithological or structural domain boundaries) using the advanced modelling proposed by \citet{bakka2019}; or we could select/deselect manually the links between adjacent SUs (Figure~\ref{fig:AdjMat}) to consider local physical---strong (permeable) or weak (impermeable)---barriers. However, the latter solution will be tedious to implement, and may introduce unnecessary subjectivity to the modelling. Lastly, we envision using information on the size of the landslides in each SU, a relevant information not currently used by our models.

More generally, we envision testing our proposed modelling framework in other areas, considering similar and different landslide types, and similar and different spatio-temporal environmental information. This will measure the applicability and flexibility of the modelling framework in different physiographic and climatic settings. 
As an example, where a multi-temporal landslide inventory is available for a large area, even with a coarser temporal resolution than the multi-temporal inventory available for Collazzone, we envision using covariates describing the spatio-temporal evolution of precipitation (\eg, rainfall totals, rainfall duration, rainfall intensity, number of rainy days) to establish a complex functional link between the medium to long term evolution of the precipitation characteristics, and the occurrence (or lack of occurrence) of landslides. We expect this to improve the currently limited ability to understand landslides in the changing climate, and to provide better climate-driven landslide projections \citep{gariano-LandslidesChanging-2016}. 
Similarly, we foresee the possibility to test the modelling framework in areas where landslides are caused by repeated geophysical (\eg, earthquake) and severe meteorological (\eg, typhoons) triggers. Where event inventory maps can be prepared after each main triggering event, which is now feasible over large areas with the existing remote sensing and image processing technologies \citep{guzzetti2012}, we expect this to improve our ability to model the evolution of complex landscapes dominated by mass-wasting processes under multiple geophysical and weather forcing \citep{burbank-DecouplingErosion-2003,Dadson-LinksErosion-2003,lave-DenudationProcesses-2004,gabet2007,larsen2010,booth2013}.

\section{Conclusions}
\label{sec:Conclusions}

We proposed a novel Bayesian modelling framework for the spatio-temporal prediction of landslides. The framework exploits a Log-Gaussian Cox Process (LGCP), which assumes that individual landslides in an area are the result of a stochastic point process driven by an unknown intensity function. We tested the modelling framework in the Collazzone area, Umbria, Central Italy, for which a detailed multi-temporal landslide inventory covering the period from before 1941 to 2014, and lithological and bedding data are available.  We used this complex space-time geomorphological and geological information to prepare five statistical models of increasing complexity. Our ``baseline" model (Mod1) solely relies on the information carried by morphometric and thematic properties, and does not account for the relative influence of spatial and temporal clustering of the landslide process. The second model (Mod2) is similar, but it allows for time-interval-specific regression constants. The next two models are more complex, and account for spatial (Mod3) and temporal (Mod4) latent effects. Lastly, our model Mod5 jointly accounts for latent temporal effects between consecutive inventories and latent spatial effects between adjacent SUs. We maintain that our most complex model Mod5 fulfils the definition of landslide hazard given in the literature. Quantification of the spatial and the temporal predictive performances of the five models revealed that our most advanced Mod5 performed better than the others model. We concluded that Mod5 is our best model, and we selected it to generate a combined landslide intensity--susceptibility classification for our study area, providing more information than traditional susceptibility zonations for land planning and management. 

Based on the results of our complex modelling experiment, we draw the following general conclusions.

\begin{itemize}

\item The landslide intensity framework introduced by \citet{Lombardo.etal:2018,lombardo2019geostatistical,lombardo2019numerical} for spatial predictions, and extended in this work for time and space-time domains, performs well, and it fulfils the requirements of the standard definition of landslide hazard within a single model. This is a significant advancement over previous landslide hazard modelling frameworks \citep{guzzetti2005probabilistic,Guzzetti2006a}. 

\item For regional geomorphological evaluations or hazard assessments, individual landslides can be considered as ``point" events, both in space and time, and a Log-Gaussian Cox Process (LGCP), or a similar model, is fully adequate for the statistical modelling of the spatial and temporal evolution of landslides in landscapes dominated by mass-wasting processes.

\item Our experiment proves that latent or ``unobserved" effects exist and they control the spatio-temporal distribution of landslides. Considering these latent space-time landslide dependencies significantly improves the model predictive performance, compared to simpler models that neglect the space-time structure of the landslide process.

\item The main limitation for complex, space-time landslide modelling resides in the availability of accurate data, and chiefly of detailed multi-temporal landslide inventories, and not in the availability of complex statistical modelling tools, which are available, or in the computational requirements, which can be relatively easily fulfilled in typical applications. This consideration should guide those interested in space-time predictive modelling of landslides in the allocation of their research time and their resource investments \citep{guzzetti1999landslidehazard,guzzetti2012}.

\end{itemize}

We expect our novel approach to the spatio-temporal prediction of landslides to enhance the ability to evaluate landslide hazard and its temporal and spatial variations, to lead to better projections of future landslides, and to improve our collective understanding of the evolution of landscapes dominated by mass-wasting processes under geophysical and weather drivers. To promote reproducible analysis and replicable experiments in different geomorphological contexts, we share as Supplementary Material the dataset, the adjacency matrix and the \texttt{R} code used in this study.

\section*{Acknowledgement}

We thank H\r{a}vard Rue, the main developer of the \texttt{R-INLA} project, and Haakon Bakka for the continuous discussions and technical support through the initial stage of this research. We are grateful to the CNR IRPI geomorphology research group (http://geomorphology.irpi.cnr.it) who have provided and updated the multi-temporal landslide inventory for the Collazzone area. Without their enduring effort, this work would not have been possible.

\section*{Appendix 1: Variables, symbols, and acronyms}

Here, we list the variables, symbols and acronyms used in the text. 

\begin{table}[!htbp]
\centering
\label{tab:appendix-1}
\begin{tabular}{lll}
\hline
Variable & Units & Explanation \\ \hline
$\beta$ & & Regression coefficient \\
$\varepsilon$ & & Innovation term in the definition of random effects \\
$\kappa$ & & Unconditional precision parameter of random effects \\
$\lambda$ & & Landslide intensity \\
$\tau$ & & Conditional precision parameter of random effects \\
$\Lambda$ &  \# & Integrated intensity, \ie, expected landslide count \\
$\hat{\Lambda}$ & \# & Estimated intensity, \ie, estimated landslide count \\
$A$ & $m^2$ & Surface area of a SU   \\
$A_L$ & $m^2$ & Surface area of a single landslide   \\
$A_{\text{LT}}$ & $m^2$ & Total landslide surface area  \\
$N$ & \# & Number of landslides in each SU   \\
$S$ & - & Susceptibility   \\
$\mathcal{N}$ & & Normal distribution \\
$W$ & - & Spatial/Temporal/Spatio-temporal random effect   \\

\vspace{-0.35cm}
& & \\ \hline
Symbol & & Explanation   \\ \hline
$\mu$ & & Mean \\
$Mo$ & & Mode \\
$\sigma$ & & Standard deviation \\
$sd$ & & Standard deviation \\
$s$ & & Space   \\
$t$ & & Time   \\

\vspace{-0.35cm}
& & \\ \hline
Acronym & & Explanation   \\ \hline
AUC & & Area Under the Curve  \\
CV & & Cross-Validation  \\
DEM & & Digital Elevation Model   \\
GSD & & Ground Sampling Distance   \\
IR & & Intensity Ratio  \\
LGCP & & Log-Gaussian Cox Process  \\
LPS & & Leica Photogrammetry Suite \\
LSE & & Latent Spatial Effect \\
LTE & & Latent Temporal Effect \\
LSTE & & Latent Spatial and Temporal Effect \\
PC & & Penalised Complexity  \\
INLA & &  Integrated  Nested  Laplace  Approximation  \\
ROC & & Receiver Operating Characteristic  \\
RSP & & Relative Slope Position  \\
SR & & Susceptibility Ratio  \\
SU & & Slope Unit  \\
TWI & & Topographic Wetness Index  \\ 
VHR & & Very High Resolution  \\ \hline
\end{tabular}
\end{table}

\newpage
\bibliographystyle{CUP}
\bibliography{landslides}
\end{document}

%% file: mylatex.tex
\def\bi{\begin{itemize}}
\def\ei{\end{itemize}}
\def\be{\begin{equation}}
\def\ee{\end{equation}}

\def\rm#1{\mathrm{#1}}

%% file: packages.tex
\usepackage[english]{babel}
\usepackage{amsfonts}
\usepackage{amssymb}
\usepackage{graphicx}
\usepackage{enumerate}
\usepackage{bm}